\documentclass[structabstract]{aa}
\usepackage[T1]{fontenc} 
\usepackage{color}
 \usepackage{ucs}
\usepackage[utf8x]{inputenc}
\usepackage{url}
\usepackage{mathrsfs}
\usepackage{amsmath} 
\usepackage{amssymb}
\usepackage{appendix}
\usepackage{multirow}
\usepackage{graphicx} 
\usepackage{varioref}
 \usepackage[T1]{fontenc}

\usepackage{float}
\usepackage{magazin_abbrev}
\usepackage{natbib}
\usepackage{sidecap}
\usepackage{color}

\def \Eqt{Eq.\thinspace}
\def \sect{Sect.\thinspace}
\def \fig{Fig.\thinspace}
\def \tab{Tab.\thinspace}
\def \App{App.\thinspace}

\begin{document}
   \title{Cosmic Shear Tomography and Efficient Data Compression using COSEBIs}


   \author{ 
   Marika Asgari
   \inst{1,}\inst{2}
          , 
          Peter Schneider
          \inst{1}
          ,  Patrick Simon
          \inst{1}
          }

   \institute{Argelander-Institut f\"ur Astronomie, Bonn University        
        \and
             SUPA, Institute for Astronomy, University of Edinburgh, Royal Observatory, Blackford Hill, Edinburgh, EH9 3HJ, U.K.
             \email{ma@roe.ac.uk}
             }


\abstract{Gravitational lensing is one of the leading tools in
  understanding the dark side of the Universe. The need for accurate,
  efficient and effective methods which are able to extract this
  information along with other cosmological parameters from cosmic
  shear data is ever growing. COSEBIs, Complete Orthogonal Sets of
  E-/B-Integrals, is a recently developed statistical measure that 
encompasses the complete E-/B-mode separable information contained in the
shear correlation functions measured on a finite angular range.}
{The aim of the present work is to test the properties of this newly
  developed statistics for a higher-dimensional parameter space and to
  generalize and test it for shear tomography.}
{We use Fisher analysis to study the effectiveness of COSEBIs. We show
  our results in terms of figure-of-merit quantities, based on Fisher
  matrices.}
{We find that a relatively small number of COSEBIs modes is always
  enough to saturate to the maximum information level.  This number is
  always smaller for `logarithmic COSEBIs' than for `linear COSEBIs',
  and also depends on the number of redshift bins, the number and
  choice of cosmological parameters, as well as the survey
  characteristics.}
{COSEBIs provide a very compact way of analyzing cosmic shear data,
  i.e., all the E-/B-mode separable second-order statistical
  information in the data is reduced to a small number of COSEBIs
  modes. Furthermore, with this method the arbitrariness in data
  binning is no longer an issue since the COSEBIs modes are
  discrete. Finally, the small number of modes also implies that
  covariances, and their inverse, are much more conveniently
  obtainable, e.g., from numerical simulations, than for the shear
  correlation functions themselves.}


\titlerunning{Tomographic cosmic shear analysis with COSEBIs}
\authorrunning{M. Asgari, P. Schneider \* P. Simon}

   \maketitle
%

\section{Introduction}

As light travels through the Universe, the gravitational potential
inhomogeneities distort its path; these distortions result in sheared
galaxy images and carry invaluable information about the matter
distribution between the observer and the source. Cosmic shear
analysis is the study of the effects of large-scale structures on
light bundles (see \citealt{2001PhR...340..291B}). Consequently, it is
one of the most promising probes for understanding the Universe,
especially dark energy. The upcoming cosmic shear surveys
(e.g. Pan-STARRS\footnote{http://pan-starrs.ifa.hawaii.edu/public/},
KIDS\footnote{http://www.astro-wise.org/projects/KIDS/},
DES\footnote{http://www.darkenergysurvey.org},
LSST\footnote{http://www.lsst.org/}, and
Euclid\footnote{http://sci.esa.int/euclid/, \citealt{2011arXiv1110.3193L}})
will have better statistical precision compared to current surveys,
which means lower noise levels, larger fields of view, deeper images,
and more accurate redshift estimations. Trustworthy and accurate
methods are able to extract all the potential information in these
future observations and make the effort put into launching them
worthwhile.

The most direct second-order statistical measurement from any weak
lensing survey are the shear two-point correlation functions
$\xi_\pm(\vartheta)$, which in reality can be determined only on a
finite interval $\vartheta_{\rm min}\le \vartheta \le \vartheta_{\rm
  max}$. These, however, cannot be used for a comparison with
theoretical models, since the shear field is in general composed of
two modes: B-modes cannot be due to leading-order lensing effects,
although they provide a measure of other effects such as shape
measurement errors and intrinsic alignment effects (see
\citealt{2010arXiv1009.2024J}; also \citealt{1998MNRAS.296..873S} and
\citealt{2002A&A...389..729S} for other effects). On the other hand,
E-modes are the only relevant modes when it comes to comparing the
cosmic shear data with models. 

Almost all of the recent analysis of cosmic shear data employ methods
of E-/B-mode separation (e.g. \citealt{2007MNRAS.381..702B} and
\citealt{2008A&A...479....9F}). These studies are done in either
Fourier or real space. For Fourier space analysis one has to find an
estimate of the power spectrum, which is sensitive to gaps and holes
in the survey and in general the survey geometry, which complicates
such analysis. On the other hand the studies in real space do not
share the same complications, since estimators of the shear
correlation functions are unaffected by such gaps.  Most of these
studies use the aperture mass dispersion
(\citealt{1998MNRAS.296..873S}), which applies compensated circular
filters to the shear field. As was shown in \cite{2002ApJ...568...20C}
and \cite{2002A&A...396....1S}, the aperture statistics, in principle,
cleanly separates the shear two-point correlations (2PCFs) into
E-/B-mode contributions. Furthermore, in the two papers just
mentioned, a decomposition of the shear 2PCFs into E- and B-mode
correlation function $\xi_{\rm E,B}(\vartheta)$ has been derived,
which also has been employed in cosmic shear analyses of survey data
(\citealt{2011arXiv1111.6622L}).

However, both the aperture statistics and the E-/B-mode correlation
functions are unobservable in practice. The aperture mass dispersion requires shape measurements of galaxy
pairs down to arbitrarily small angular scales. Since this is not
feasible in real data, usually ray-tracing simulations fill in the
gap, resulting in biases and E-/B-mode mixing (see
\citealt{2006A&A...457...15K}). On the other hand, the determination
of $\xi_{\rm E,B}(\vartheta)$ requires the knowledge of
$\xi_-(\vartheta')$ out to infinite $\vartheta'$. Hence, in both
cases, determining E-/B-mode separated statistics requires some sort of
data invention.


To overcome these problems,
\cite{2007A&A...462..841S} derived general conditions and relations
for $E$-/$B$-statistics based upon two-point statistical quantities,
namely 2PCFs and convergence power spectra. They defined the quantities 
\begin{align}
\label{E-B}
E = \frac{1}{2} \int_0^{\infty}\mathrm{d}\vartheta\:\vartheta\:[T_+(\vartheta)\xi_+(\vartheta) + T_-(\vartheta)\xi_-(\vartheta)]\;,\\
B = \frac{1}{2} \int_0^{\infty}\mathrm{d}\vartheta\:\vartheta\:[T_+(\vartheta)\xi_+(\vartheta) - T_-(\vartheta)\xi_-(\vartheta)]\;;
\end{align} 
provided that 
the filter functions satisfy
\begin{equation}
\label{T}
\int_0^{\infty}\mathrm{d}\vartheta\:\vartheta\:T_+(\vartheta) \mathrm{J}_0(\ell\vartheta)=
\int_0^{\infty}\mathrm{d}\vartheta\:\vartheta\:T_-(\vartheta) \mathrm{J}_4(\ell\vartheta)\;,
\end{equation} 
$E$ depends only on the E-mode shear, and $B$ depends only on the
B-mode shear (with the aperture dispersion being one particular example).
Moreover, they have shown that in order to obtain these statistics
from the shear 2PCFs on a
finite angular interval,
$0<\vartheta_\mathrm{min}<\vartheta<\vartheta_\mathrm{max}<\infty$,
the filter function $T_+$ should have finite support on the same
angular interval and satisfy
\begin{flalign}
\label{T+cond}
\int_{\vartheta_{\mathrm{min}}}^{\vartheta_{\mathrm{max}}}\mathrm{d}\vartheta\:\vartheta\:T_+(\vartheta) & =0= \int_{\vartheta_{\mathrm{min}}}^{\vartheta_{\mathrm{max}}}\mathrm{d}\vartheta\:\vartheta^3\:T_+(\vartheta)\;.
\end{flalign}

Whereas all solutions to the above relations provide statistics which
cleanly separate E-/B-modes on a finite interval, different solutions
may vary in their information contents. For example, the ring
statistics introduced in \cite{2007A&A...462..841S} has a lower
signal-to-noise for a fixed angular range than the aperture
dispersion, which, however, is compensated by its more diagonal noise-covariance
matrix resulting in comparable Fisher matrices with aperture mass
dispersion (\citealt{2010MNRAS.401.1264F}).

Recently, a complete solution of this issue was obtained
(\citealt{2010A&A...520A.116S}, hereafter SEK) by defining Complete
Orthogonal Sets of E-/B-Integrals (COSEBIs). COSEBIs
capture the {\it full} information of the shear 2PCFs on a finite
interval {\it which is E-/B-mode separable}. In fact, 
SEK have shown that a small number of COSEBIs contain all the information
about the cosmological dependence in their two-parameter
model. Furthermore, they showed that COSEBIs in fact put tighter
constraints on these parameters compared to the aperture mass
dispersion. \cite{2011MNRAS.418..536E}
obtained a similar conclusion
for a five-parameter cosmological model. Therefore, the set of COSEBIs
not only capture the full information, but also provide a highly
efficient and simple method for data compression.

In this paper we further generalize the analysis in SEK to seven
cosmological parameters, $\sigma_8$, $\Omega_\mathrm{m}$,
$\Omega_\Lambda$, $w_0$, $n_\mathrm{s}$, $h$, and $\Omega_\mathrm{b}$,
and investigate the effect of tomography on the results. Tomography,
the joint analysis of shear auto- and cross-2PCFs of galaxy
populations with different redshift distributions, is a powerful tool
for cosmological analysis (\citealt{2006astro.ph..9591A};
\citealt{2006ewg3.rept.....P}), in particular in multi-dimensional
parameter space (see \citealt{2010A&A...516A..63S} for a recent paper
on constraints on dark energy from cosmic shear analysis with
tomography). We use Fisher analysis throughout our paper to represent
the constraining power of COSEBIs, and compare the results from a
medium-sized with that of a large cosmic shear survey.

In \sect$\ref{COSEBIs}$ we summarize the method used in SEK and write
the corresponding relations for shear tomography. In
\sect$\ref{CosmoModel}$ we briefly explain our choice of cosmology,
and in \sect$\ref{Covari}$ the covariance of COSEBIs is shown. We
present our figure-of-merit based on Fisher analysis and show the
results for the seven cosmological parameters and up to eight redshift
bins in \sect$\ref{result}$. Finally we conclude by summarizing the
most important results of the previous sections and emphasizing the
advantages of COSEBIs over other methods of cosmic shear analysis. We
have also derived an analytic solution to the linear COSEBIs weight
functions presented in \App\ref{linearCOSEBIs}.  

\section{COSEBIs}
\label{COSEBIs}

There is an infinite number of filter functions $T_+(\vartheta)$ satisfying
Eq.\thinspace $\eqref{T+cond}$. Such filters can be expanded in sets
of orthogonal functions, labeled $T_{+n}(\vartheta)$; the
corresponding $T_{-n}(\vartheta)$ are obtained from solving
\Eqt$\eqref{T}$ which can be inverted explicitly (\citealt{2002A&A...396....1S}).
Accordingly, the corresponding E/B-statistics are denoted by
$E_n$ and $B_n$, respectively. Here we will also consider the case
that different galaxy populations can be distinguished (mainly by
their redshifts); therefore, one can measure auto- and
cross-correlations functions of the shear,
$\xi^{ij}_\pm(\vartheta)$. We denote the corresponding COSEBIs by
$E^{ij}_n$ and $B^{ij}_n$. They are related to the auto- and
cross-power spectra of the convergence, by
\begin{align}
\label{EnofWn}
E_n^{ij} = \int_0^{\infty} \frac{\mathrm{d}\ell\:\ell\:}{2\pi}P^{ij}_{\mathrm{E}}(\ell)W_n(\ell)\;,\\
\label{BnofWn}
B_n^{ij} =  \int_0^{\infty} \frac{\mathrm{d}\ell\:\ell\:}{2\pi}P^{ij}_{\mathrm{B}}(\ell)W_n(\ell)\;,
\end{align} 
where $P^{ij}_\mathrm{E/B}$ are the E-/B-cross convergence power spectra of galaxy
populations $i$ and $j$ (see \citealt{2002A&A...396....1S}), and are
related to the 2PCFs by
\begin{align}
\label{xi+}
\xi_+^{ij}(\vartheta)=\int_0^{\infty} \frac{\mathrm{d}\ell\:\ell\:}{2\pi} \mathrm{J}_0(\ell\vartheta)[P_{\mathrm{E}}^{ij}(\ell)+P_{\mathrm{B}}^{ij}(\ell)]\;,\\
\label{xi-}
\xi_-^{ij}(\vartheta)=\int_0^{\infty} \frac{\mathrm{d}\ell\:\ell\:}{2\pi} \mathrm{J}_4(\ell\vartheta)[P_{\mathrm{E}}^{ij}(\ell)-P_{\mathrm{B}}^{ij}(\ell)]\;.
\end{align}
Inserting the above relations into \Eqt$\eqref{E-B}$, one can find 
relations connecting $W_n$ to $T_{\pm n}$
\begin{align}
\label{Wn}
W_n(\ell) & =  \int_{\vartheta_{\mathrm{min}}}^{\vartheta_{\mathrm{max}}}\mathrm{d}\vartheta\:\vartheta\:T_{+n}(\vartheta) \mathrm{J}_0(\ell\vartheta) \\ 
& = \int_{\vartheta_{\mathrm{min}}}^{\vartheta_{\mathrm{max}}}\mathrm{d}\vartheta\:\vartheta\:T_{-n} (\vartheta) \mathrm{J}_4(\ell\vartheta)\;.
\end{align}

Any type of cosmic shear analysis needs some sort of error assessment. In particular Fisher analysis, used in the present work, depends on the noise-covariance of the statistics employed. The noise-covariance of COSEBIs for several galaxy populations assuming Gaussian shear fields (see \citealt{2008A&A...477...43J}) is

\begin{align}
\label{CmnP}
C_{mn}^{X(ij,kl)} & \equiv  \langle X^{ij}_m X^{kl}_n\rangle-\langle X^{ij}_m\rangle\langle X^{kl}_n\rangle \nonumber \\
& = \frac{1}{2 \pi A}\int_0^{\infty} \mathrm{d}\ell\:\ell\:W_m(\ell)W_n(\ell)\nonumber \\
&\times \left( \bar{P}^{ik}_{\mathrm{X}}(\ell)\bar{P}^{jl}_{\mathrm{X}}(\ell)+\bar{P}^{il}_{\mathrm{X}}(\ell)\bar{P}^{jk}_{\mathrm{X}}(\ell)\right)\;,
\end{align} where 

\begin{equation}
\bar{P}^{ik}_{\mathrm{X}}(\ell):= P^{ik}_{\mathrm{X}}(\ell)+\delta_{ik}\frac{\sigma_{\epsilon}^2}{2\bar{n}_i}\;,
\end{equation} 
and X stands for either E or B. The survey parameters are also
included in \Eqt\eqref{CmnP} with the survey area, $A$, the galaxy
intrinsic r.m.s ellipticity, $\sigma_{\epsilon}$, and the mean number
density of galaxies in each redshift bin, $\bar{n}_i$. 

In a recent paper, \cite{2011ApJ...734...76S} have shown that the Gaussian
covariance model in \cite{2008A&A...477...43J} overestimates the
true Gaussian covariance for surveys with small area ($ A \lesssim
1000 \:\mathrm{deg}^2$), and they have developed a fitting formula to
correct for this discrepancy; in spite of their findings we will stick
to the estimation of \cite{2008A&A...477...43J}, since the fitting
formula in the latter paper depends on source redshift and is
developed for a single source galaxy redshift, making it
non-applicable for this work.

Alternatively, one can write the covariance (\Eqt$\ref{CmnP}$) in terms of $T_{\pm
  n}$ and the two-point correlation functions' covariance (see
SEK). However, in this approach double integrals over the covariance
of 2PCFs slow down the calculations.


\subsection{The COSEBIs filter and weight functions}
\label{filters}
SEK constructed two complete orthogonal sets of functions, linear
and logarithmic COSEBIs (hereafter Lin- and Log-COSEBIs respectively),
by considering \Eqt\eqref{T+cond}, and imposing orthogonality
conditions on the $T_{+n}$ filters. Once the
$T_{+n}$ filters are known, the $T_{-n}$ filters can be calculated via
\Eqt\eqref{T}. The Lin-COSEBIs filters are polynomials in $\vartheta$,
the angular separation of galaxies, while the Log-COSEBIs filters are polynomials
in $\ln(\vartheta)$.

The output of theoretical cosmological models which is of relevence
here is the power spectrum. Hence, the quickest way to treat COSEBIs in theory is to work in $\ell$-space
and to use \Eqt$\eqref{CmnP}$ for the covariance, without taking the
detour of calculating the shear 2PCFs and their covariance. As a result we need to calculate the
$W_n(\ell)$ functions which are the Hankel transform (\Eqt$\ref{Wn}$) of
their real-space counterparts, $T_{\pm n}$. For convenience, we choose to evaluate $W_n(\ell)$ from their integral relation with $\mathrm{J}_0$ and $T_{+n}$.
Since both $\mathrm{J}_0$ and $T_{+n}$ are oscillating functions, evaluating these integrals is rather
challenging, in particular for large $\ell$.  A piece-wise
integration, from one extremum to the next, is used in the present
work to evaluate $W_n(\ell)$. \App\ref{linearCOSEBIs} contains more
details about the numerical integrations and also a (semi-)analytic
formula for the linear $W_n$ functions.

\begin{figure*}
  \begin{center}
    \begin{tabular}{c}
      \resizebox{150mm}{!}{\includegraphics{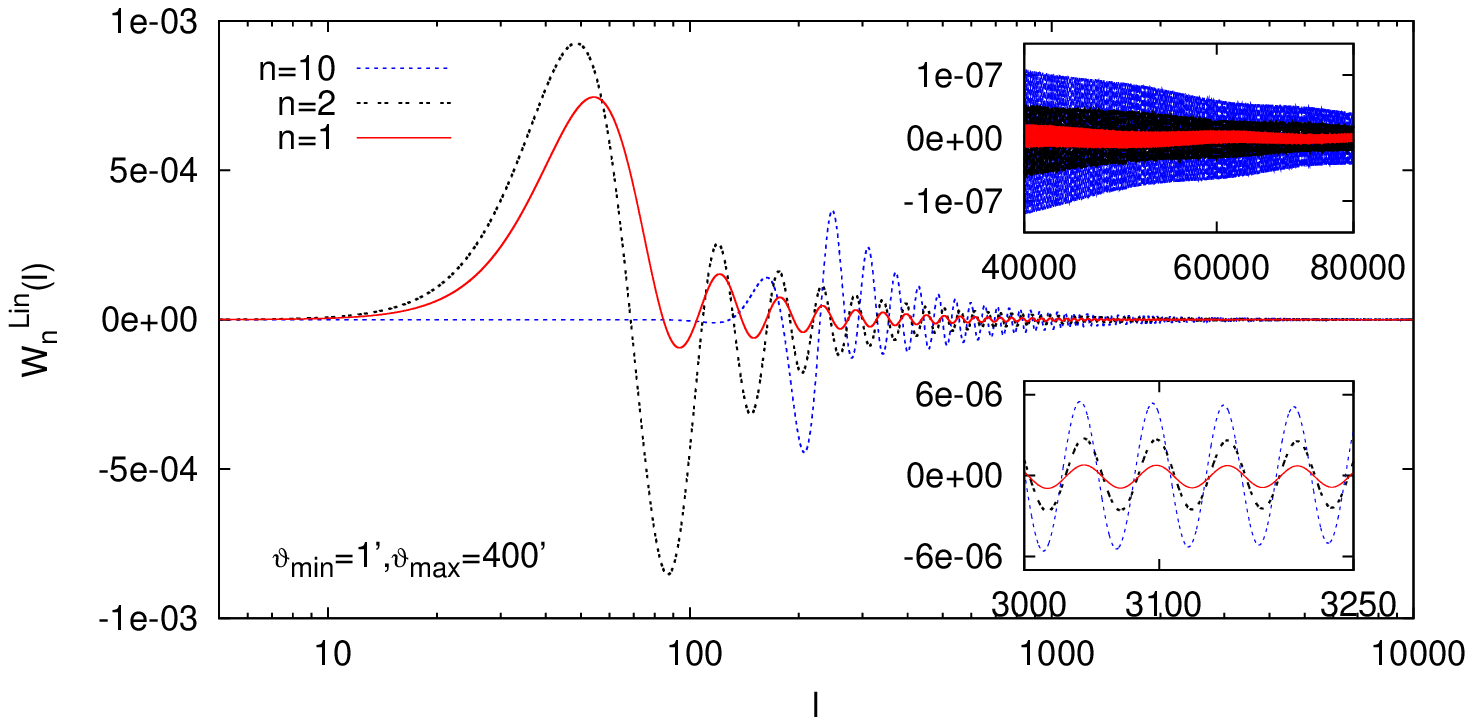}} \\
       \resizebox{150mm}{!}{\includegraphics{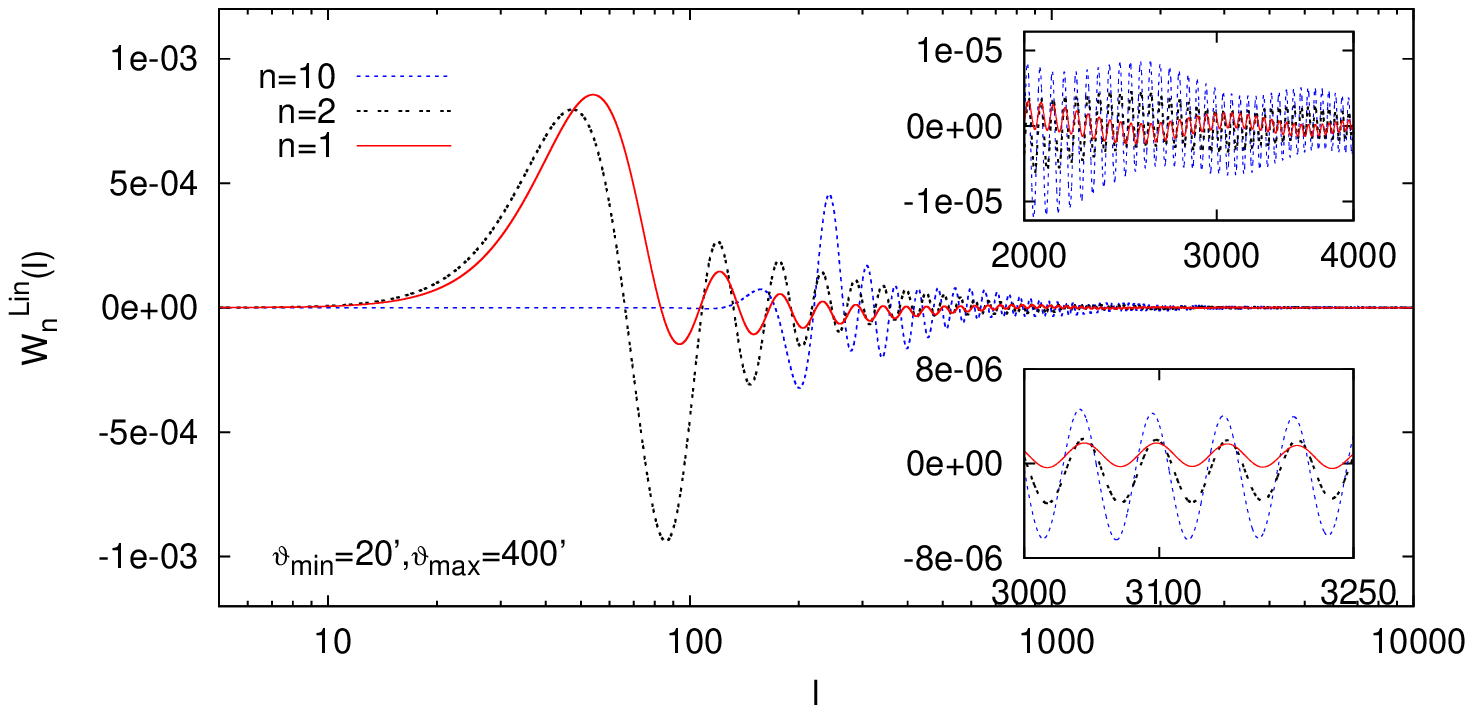}} \\
      \resizebox{150mm}{!}{\includegraphics{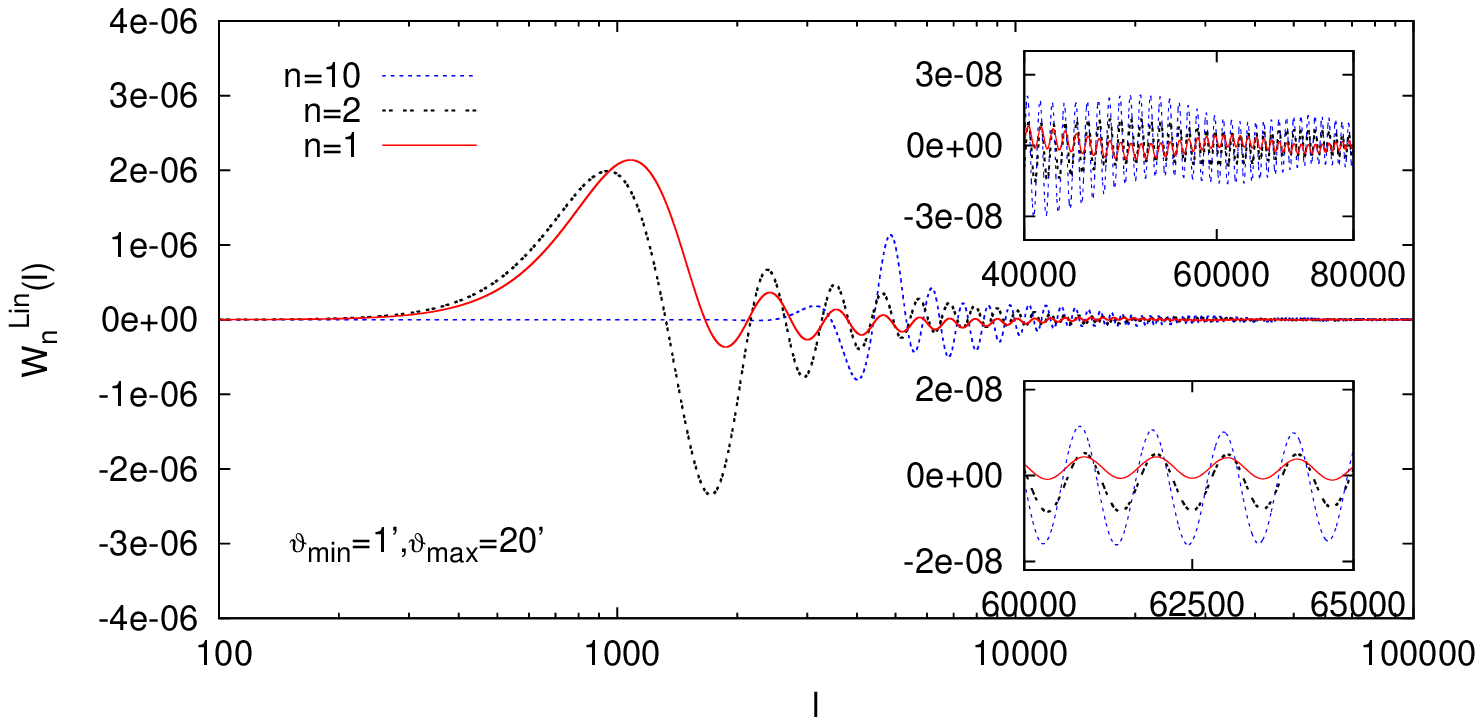}}
    \end{tabular}
    \caption{\small{The weight functions $W_n^\mathrm{Lin}(\ell)$ are the Hankel transforms of $T_{\pm}^\mathrm{Lin}(\vartheta)$ as in \Eqt$\eqref{Wn}$. In the blow-ups, the two modes of oscillation for each $W_n^\mathrm{Lin}$ can be seen, the lower frequency mode and the higher frequency mode which are inversely proportional to $\vartheta_{\mathrm{min}}$ and $\vartheta_{\mathrm{max}}$, respectively. The overall amplitude of the oscillations strongly depends on $n$ and $\vartheta_{\mathrm{max}}$.} }

    \label{Wn-plot}
  \end{center}
\end{figure*}

As is explained in SEK, the Log-COSEBIs are more efficient for a cosmic
shear analysis. The reason is that unlike the linear filter functions
which oscillate fairly uniformly in linear scale, the logarithmic
$T_{+n}^\mathrm{Log}$ have their roots fairly uniformly distributed in
$\log(\vartheta)$, i.e., they are more sensitive to variations of
$\xi_\pm$ on smaller scales. Combining this property with the fact
that most of the cosmic shear information is contained in these
smaller scales shows that it is more reasonable to employ Log-COSEBIs. In the next section we will show the difference of the Log-
and Lin-COSEBIs using our figure-of-merit.

\begin{figure*}
  \begin{center}
    \begin{tabular}{c}
      \resizebox{150mm}{!}{\includegraphics{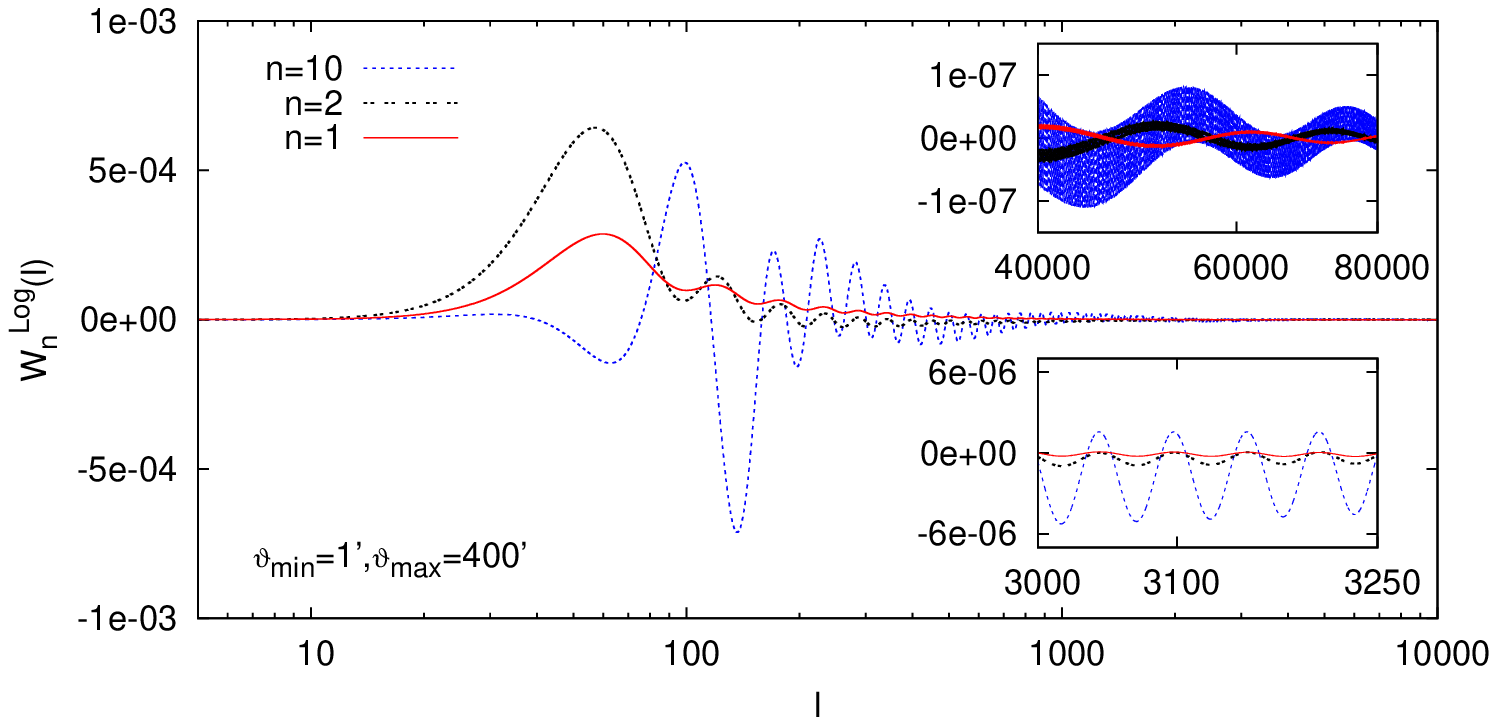}} \\
        \resizebox{150mm}{!}{\includegraphics{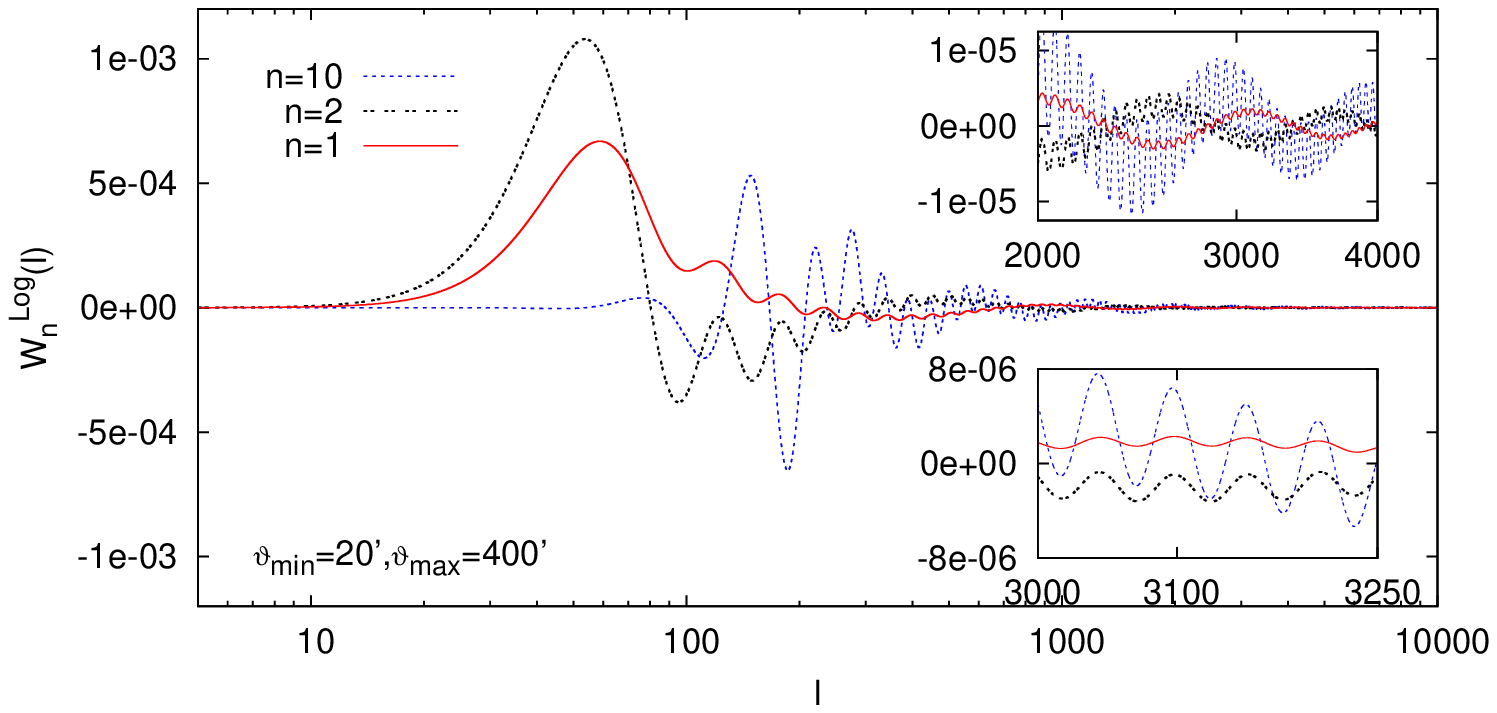}} \\
      \resizebox{150mm}{!}{\includegraphics{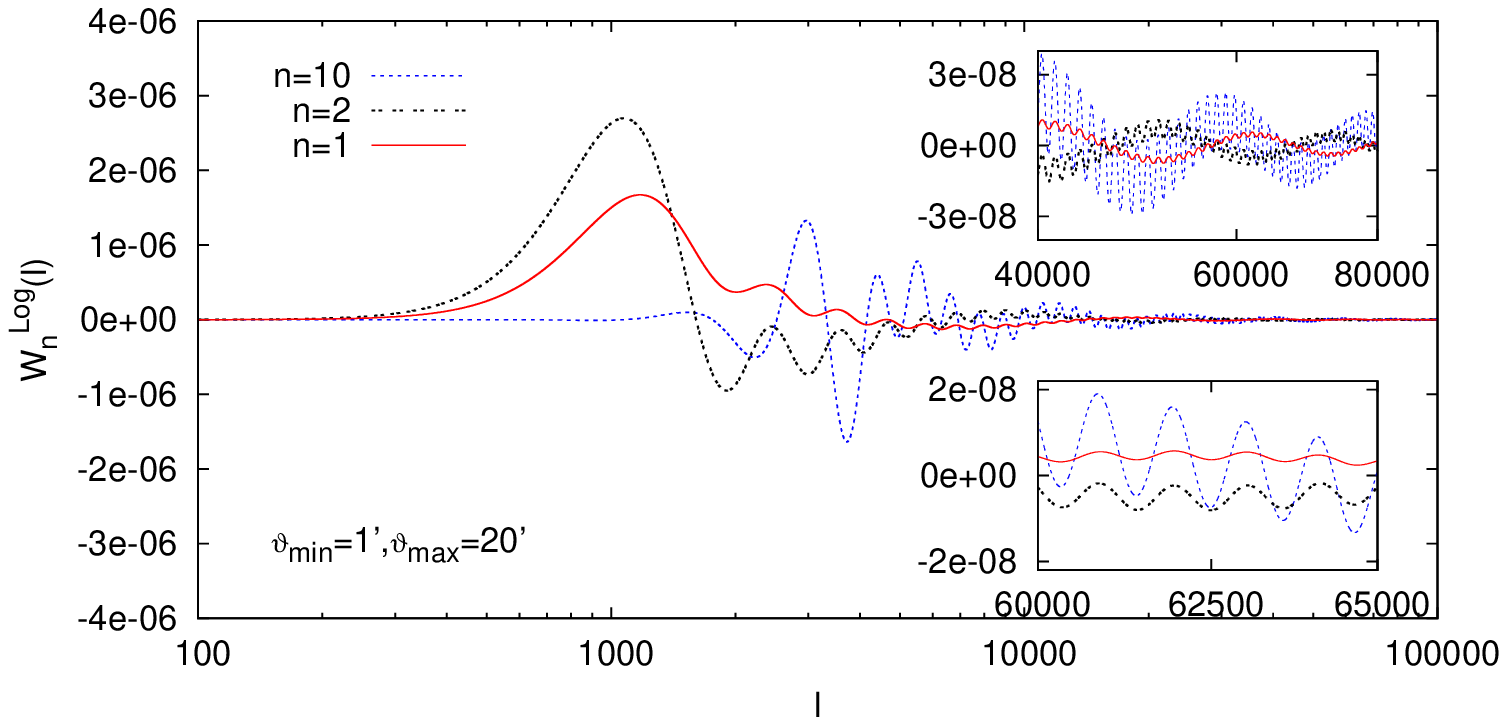}}
    \end{tabular}
    \caption{\small{The weight functions $W_n^\mathrm{Log}(\ell)$ are
        the Hankel transformation of $T_{\pm}^\mathrm{Log}(\vartheta)$
        as in \Eqt$\eqref{Wn}$. Similar to the $W_n^\mathrm{Lin}$,
        the position of the first peak depends mainly on
        $\vartheta_\mathrm{max}$ and is rather insensitive to
        $\vartheta_\mathrm{min}$. The difference between the two sets
        of linear and logarithmic function can be seen most
        prominently in the blow-ups; the lower frequency oscillations
        are more pronounced in this case.}} 
    \label{Wnlog-plot}
  \end{center}
\end{figure*}

In \fig\ref{Wn-plot} and \fig\ref{Wnlog-plot} the behavior of linear and logarithmic COSEBIs weight functions,
$W_n^\mathrm{Lin}(\ell)$ and $W_n^\mathrm{Log}(\ell)$, for three angular
ranges can be seen. The $W_n^\mathrm{Log}(\ell)$ and
$W_n^\mathrm{Lin}(\ell)$ have different yet similar oscillatory
properties. They both die out rapidly with increasing $\ell$ but the
lower frequency oscillations of $W_n^\mathrm{Log}(\ell)$ are more
prominent. They show approximately the same inverse relation to
$\vartheta_{\mathrm{max}}$ and $\vartheta_{\mathrm{min}}$ for their
lower and upper limits.

\section{Cosmological Model}
\label{CosmoModel}
The cosmological model assumed in the present work is a wCDM model
(\citealt{2003RvMP...75..559P} and references therein), i.e., a cold dark
matter model including a dynamical dark energy with an
equation-of-state parameter $w_0$. The fiducial value of the
parameters involved are listed in \tab\ref{cosmparam}.

The starting point in the analysis is to derive the matter power
spectrum. For the linear power spectrum we used the
\cite{1984ApJ...285L..45B} transfer function, and the halo fit formula
of \cite{2003MNRAS.341.1311S} for a fit of the non-linear regime.

\begin{figure}
  \begin{center}
    \begin{tabular}{c}
      \resizebox{85mm}{!}{\includegraphics{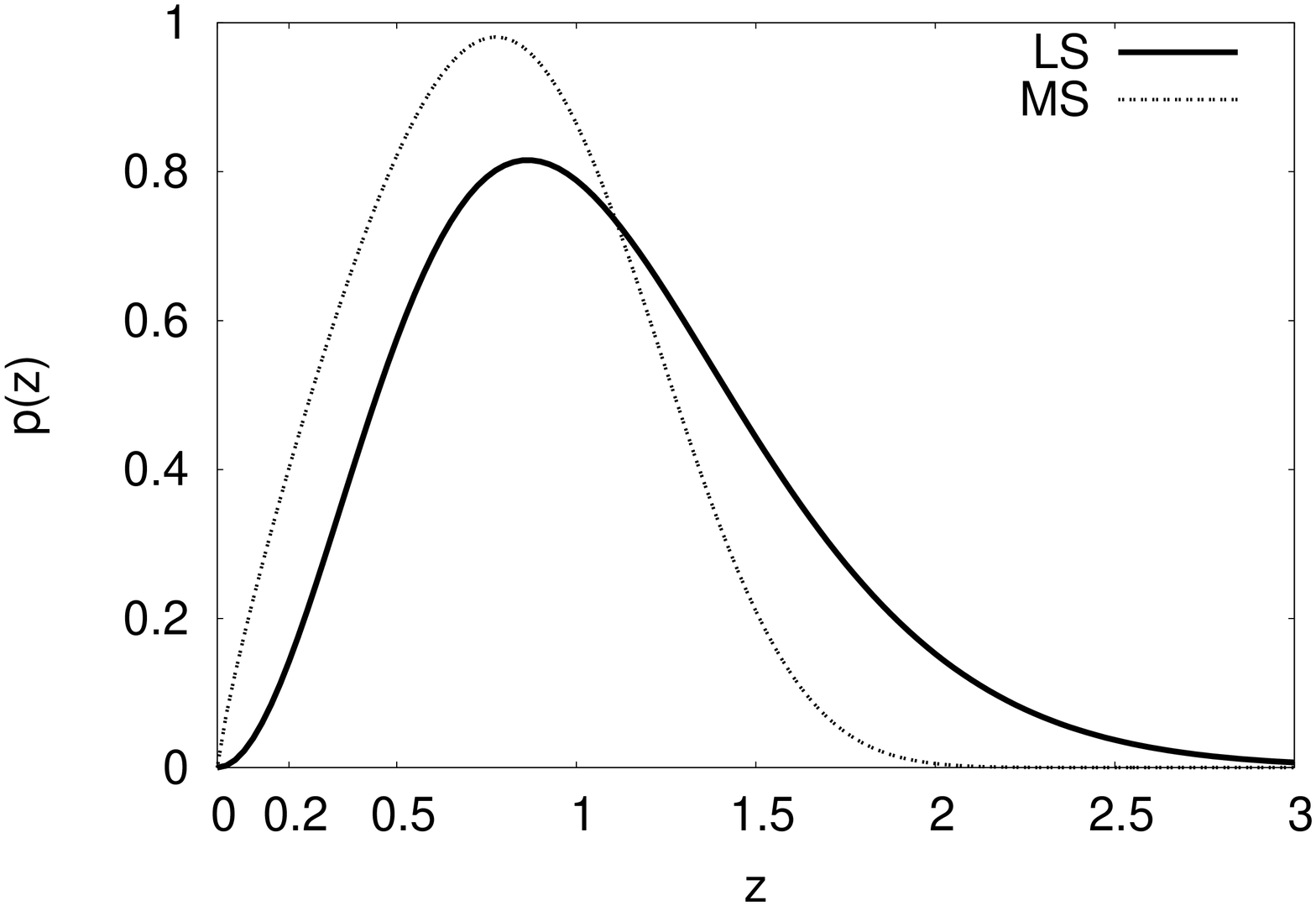}}
    \end{tabular}
    \caption{\small{The overall source redshift probability
        distribution of source galaxies assumed for the two
        surveys. LS has a deeper source distribution compared to MS.}}
    \label{pofz-Euclid-CFHTLS}
  \end{center}
\end{figure}

To calculate the convergence power spectrum we need the redshift
distribution of galaxies. The overall redshift probability distribution is parametrized by
\begin{equation}
\label{Pz}
p(z)=\frac{\beta}{z_0\: \Gamma[(1+\alpha)/\beta]}\left(\frac{z}{z_0}\right)^\alpha \exp\left[-\left(\frac{z}{z_0}\right)^\beta\right]\;,
\end{equation} 
which represents the galaxy distribution fairly well (it is a
generalization of \citealt{1996ApJ...466..623B}). The parameters,
$\alpha$, $\beta$, and $z_0$ depend on the survey. We consider a
medium and a large survey (hereafter MS and LS respectively). The MS
has the same area as the CFHTLS (\citealt{2008A&A...479....9F}), a
current survey, while the LS covers the whole extragalactic sky and
represents future surveys. The parameters of our two model surveys are
given in \tab$\ref{surveyTable}$, and the corresponding redshift
distributions are plotted in \fig$\ref{pofz-Euclid-CFHTLS}$.

\begin{table}
\caption{\small{The fiducial cosmological parameters consistent with
    WMAP 7-years results. The normalization of the power spectrum,
    $\sigma_8$, is the standard deviation of perturbations in a sphere
    of radius 8 $\:h^{-1}$Mpc today. $\Omega_\mathrm{m}$,
    $\Omega_\Lambda$, and $\Omega_\mathrm{b}$ are the matter, the dark
    energy and the baryonic matter density parameters,
    respectively. $w_0$ is the dark energy equation of state
    parameter, which is equal to the ratio of dark energy pressure to
    its density. The spectral index, $n_\mathrm{s}$, is the slope of
    the primordial power spectrum. The dimensionless Hubble constant,
    in $H_0=100 h $ km s$^{-1}$ Mpc$^{-1}$, characterizes the rate of expansion today. }} 
\begin{center}
\begin{tabular}{ | c | c | c | c | c | c | c | }
  \hline
  $\sigma_8$ & $\Omega_\mathrm{m}$ & $\Omega_\Lambda$ & $w_0$ & $n_\mathrm{s}$ & $h$ & $\Omega_\mathrm{b}$ \\
  \hline
  0.8  & 0.27  & 0.73  &  $-1.0$ &  0.97 & 0.70 & 0.045 \\
  \hline
\end{tabular}
\end{center}
\label{cosmparam}
\end{table}

\begin{table*}
\caption{\small{The redshift distribution parameters and the survey
    parameters for our medium and large surveys. $\alpha$, $\beta$,
    and $z_0$ determine the total redshift distribution of sources,
    while $z_\mathrm{min}$ and $z_\mathrm{max}$ indicate the minimum
    and the maximum redshifts of the sources considered. $A$ is
    the survey area in units of deg$^2$, $\sigma_{\epsilon}$ is the
    galaxy intrinsic ellipticity dispersion, and $\bar n$ is the mean
    number density of sources per square arcminute in the field. }} 
\begin{center}
\begin{tabular}{c|c|c|c|c|c|c|c|c|}
\cline{2-9}
&  \multicolumn{5}{|c|}{z-distribution parameters} 
&  \multicolumn{3}{|c|}{survey parameters}
\\ \cline{2-9}
&  $\alpha$ & $\beta$ & $z_0$ & $z_\mathrm{min}$ & $z_\mathrm{max}$ & $A$ & $\sigma_{\epsilon}$ & $\bar n$\\ \cline{1-9}
\multicolumn{1}{|c|}{\multirow{1}{*}{MS}} 
 &  0.836 &  3.425 & 1.171 & 0.2 &  1.5  &  170 & 0.42 & 13.3 \\ \cline{1-9}
\multicolumn{1}{|c|}{\multirow{1}{*}{LS}} 
 & 2.0 &  1.5 & 0.71 & 0.0 & 2.0 & 20000 & 0.3 & 35 \\ \cline{1-9}
\end{tabular}
\end{center}
\label{surveyTable}
\end{table*}

\begin{figure}
  \begin{center}
    \begin{tabular}{c}
      \resizebox{85mm}{!}{\includegraphics{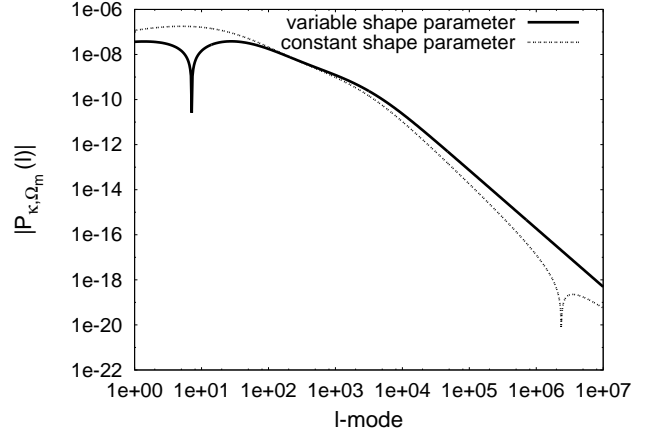}}
    \end{tabular}
    \caption{\small{The absolute value of the derivative of the convergence power spectrum with respect to $\Omega_\mathrm{m}$. Both of the curves rely on a five point stencil method where 4 nearby points have to be evaluated. The solid curve is drawn assuming all parameters are fixed except $\Omega_\mathrm{m}$ and $\Gamma$, in contrast to the dotted curve where instead of $\Gamma$, $h$ or $\Omega_\mathrm{b}$ are variable.}}
    \label{power}
  \end{center}
\end{figure}

Constructing the Fisher matrix requires the derivatives of the E-mode
COSEBIs and of their covariances with respect to the parameters. For
example, to take the derivative with respect to $\Omega_\mathrm{m}$,
its relation to the shape parameter, $\Gamma$, should be notified. In
the present work we use the 
\cite{1995ApJS..100..281S} relation, 
\begin{equation}
\label{shapeparameter}
 \Gamma=\Omega_\mathrm{m}\:h \exp[-\Omega_\mathrm{b}(1+\sqrt{2h}/\Omega_\mathrm{m})]\;.
\end{equation}
In their derivatives with respect to $\Omega_\mathrm{m}$, SEK assumed
a constant $\Gamma$, equivalent to allowing $h$ or $\Omega_\mathrm{b}$
to vary accordingly (the only dependence of the convergence power
spectrum on $h$ or $\Omega_\mathrm{b}$ comes through $\Gamma$). In the
present work $h$ and $\Omega_\mathrm{b}$ are independent parameters
and $\Gamma$ depends explicitly on $\Omega_\mathrm{m}$.  The
difference between the two approaches is not negligible, as shown in
\fig$\ref{power}$ which displays the derivative of the power spectrum
with respect to $\Omega_\mathrm{m}$ in both cases.  This difference is
due to the non-linear relation between $h$ and $\Gamma$. To justify
our choice of parametrization, we just mention that the constraints
from cosmological probes on $h$ is tighter compared to $\Gamma$, and
that makes it a more natural choice especially when priors are used.

\section{COSEBIs Covariance}
\label{Covari}
\begin{figure}
  \begin{center}
    \begin{tabular}{c}
      \resizebox{85mm}{!}{\includegraphics{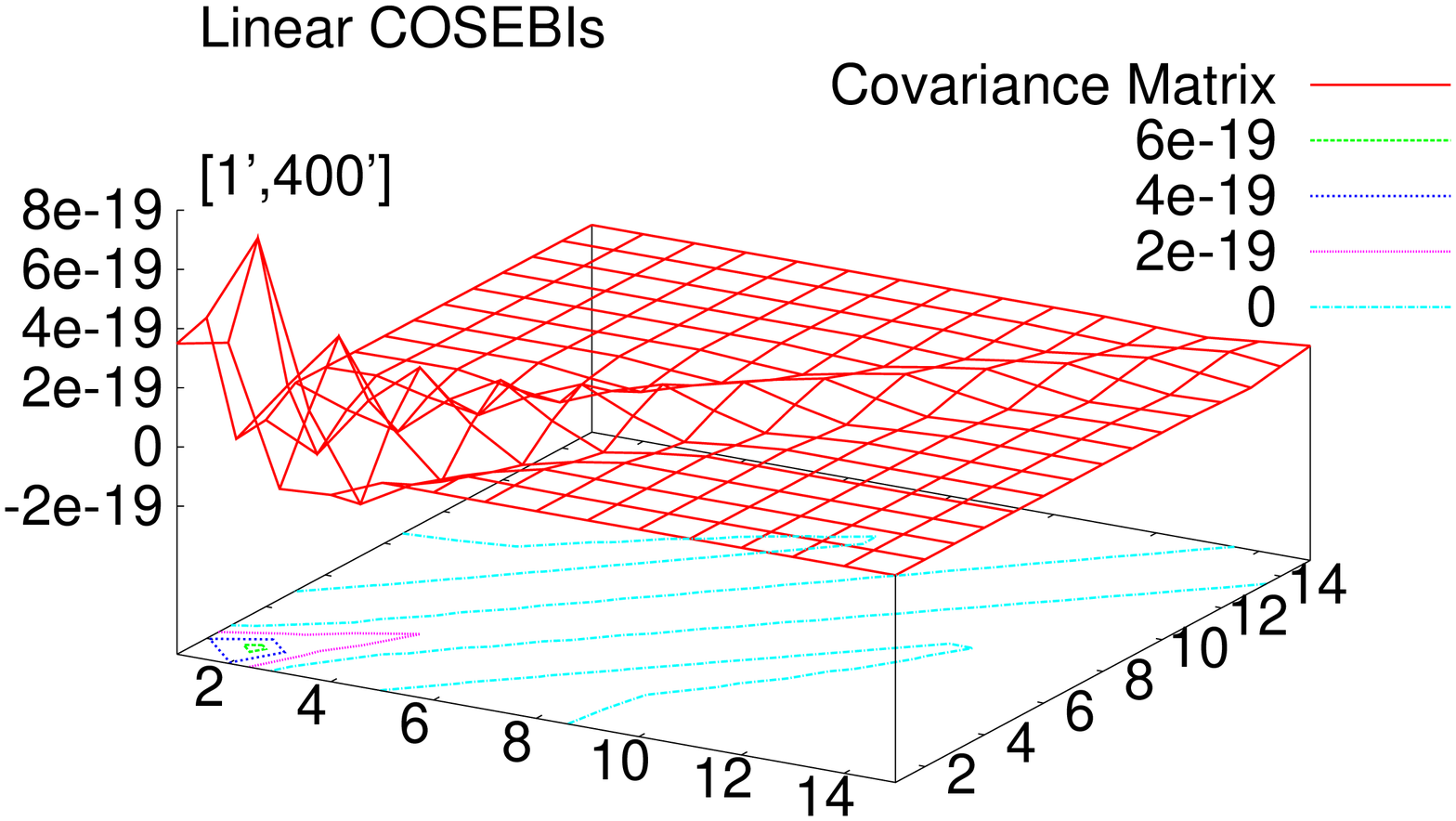}}\\
      \resizebox{85mm}{!}{\includegraphics{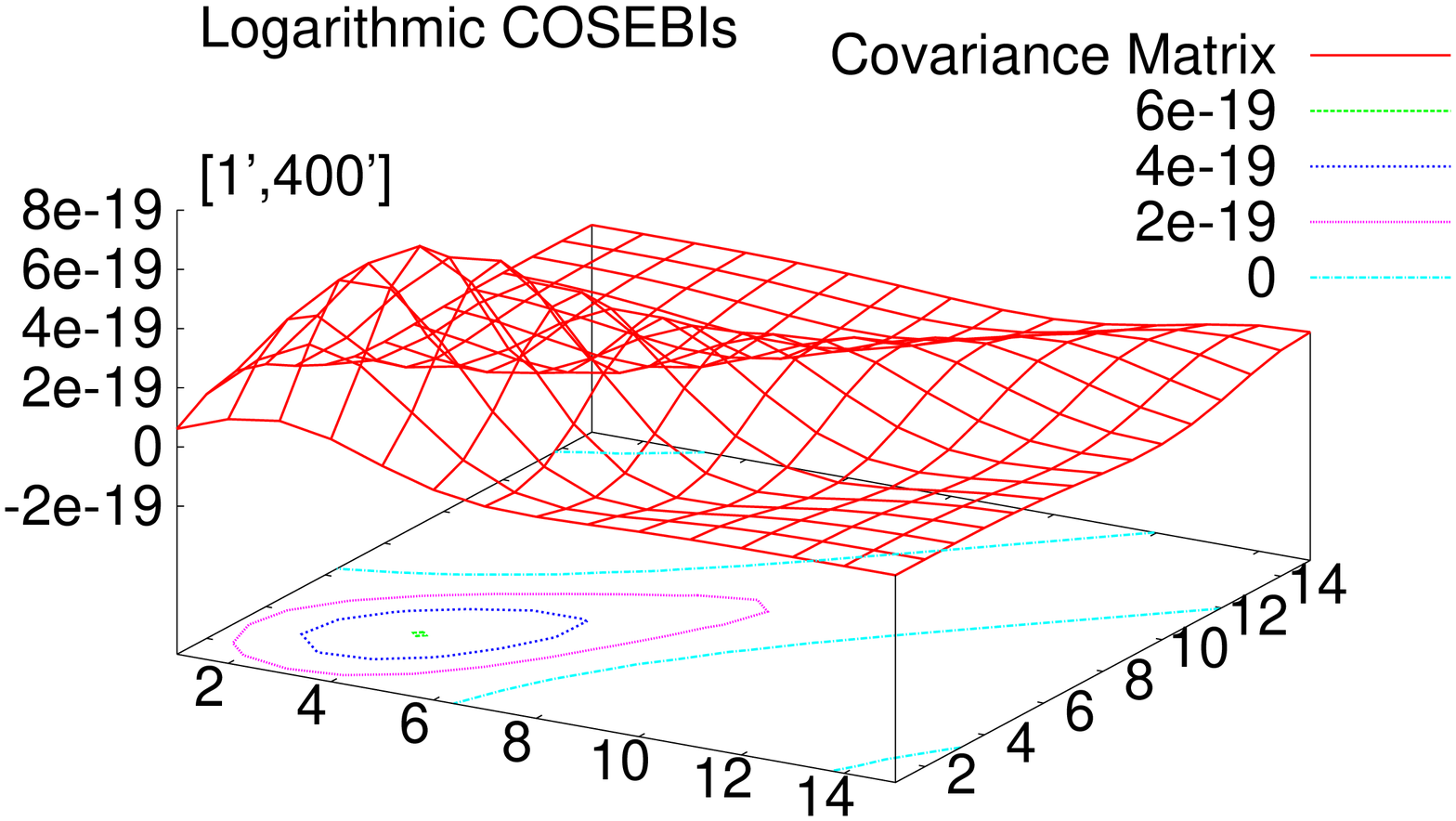}}
    \end{tabular}
    \caption{\small{A 3D representation of the non-tomographic
        covariance of 15 E-mode COSEBIs for an angular range of
        $[1',400']$, for MS parameters. The $x$- and $y$- axes
        correspond to the elements of the covariance matrix, and the
        value of the vertical axis shows the value of the covariance
        of the corresponding element. A contour representation of the
        covariance is shown for each plot at its base.}} 
    \label{covariance}
  \end{center}
\end{figure}

\begin{figure}
  \begin{center}
    \begin{tabular}{c}
      \resizebox{85mm}{!}{\includegraphics{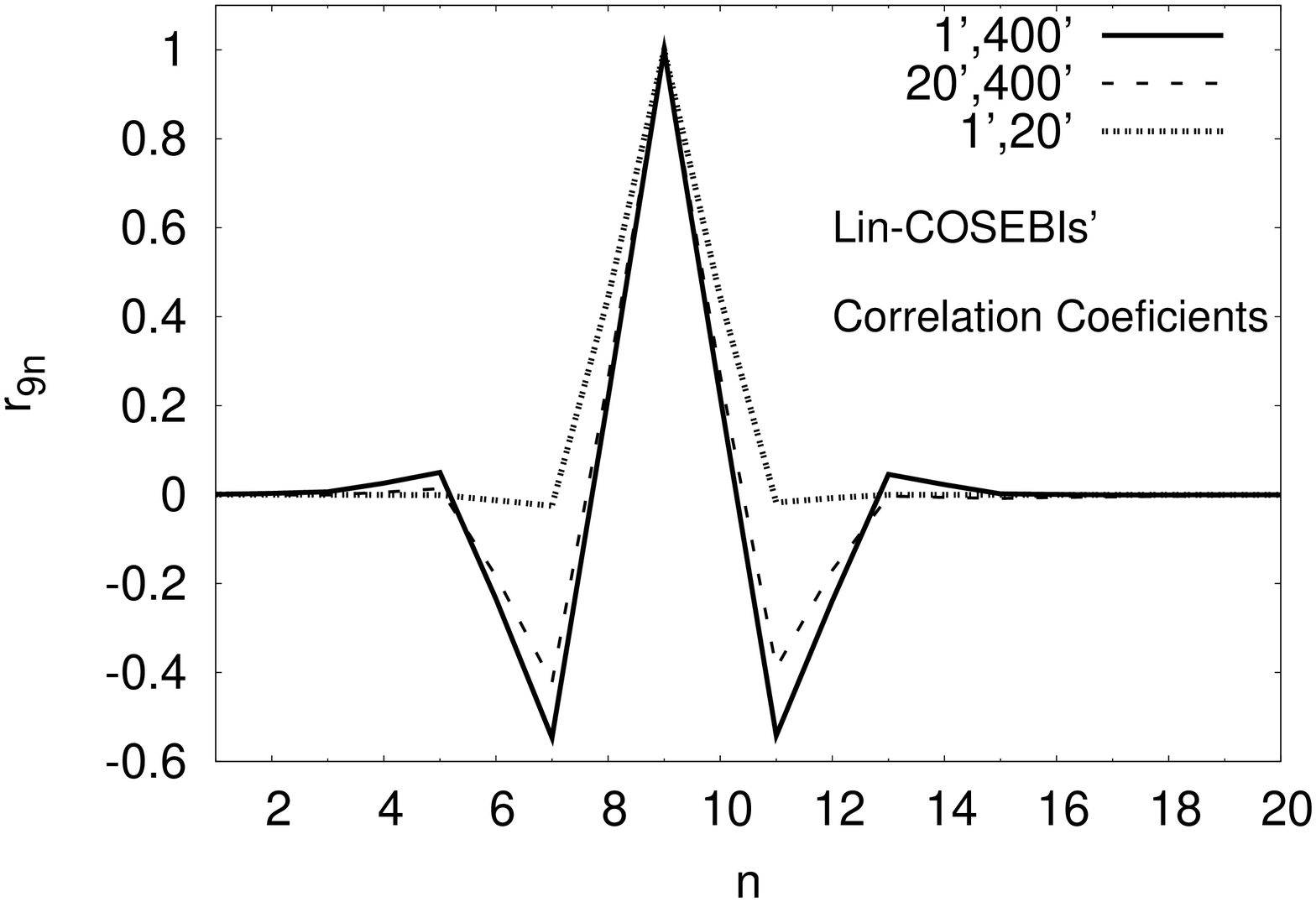}}\\
      \resizebox{85mm}{!}{\includegraphics{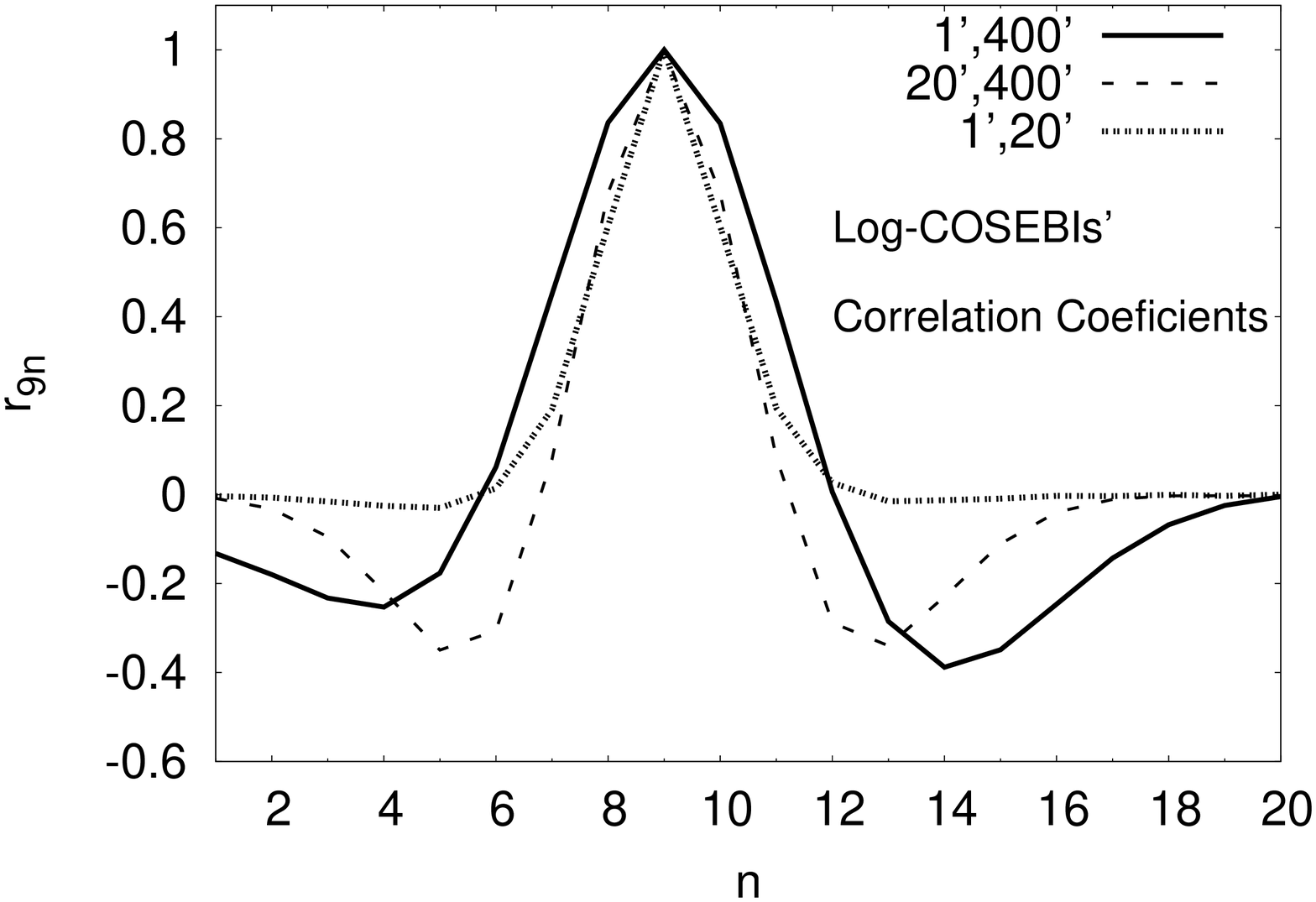}}
    \end{tabular}
    \caption{\small{The correlation coefficients of non-tomographic
        COSEBIs for different angular ranges
        $[\vartheta_{\rm min},\vartheta_{\rm max}]$
at $m=9$, for the MS parameters. Here $M$, the capital subscripts, are
equal to the COSEBIs mode, $m$.}} 
    \label{RMT3theta}
  \end{center}
\end{figure}

\fig$\ref{covariance}$ shows the noise-covariance of linear and logarithmic
E-mode COSEBIs for the model parameters of the MS (the covariance has
a similar behavior in the case of the LS but with a different
amplitude). This covariance is calculated from \Eqt\eqref{CmnP}
assuming a single source redshift distribution (\Eqt$\ref{Pz}$). Moreover, by
defining the correlation coefficients of COSEBIs, 
\begin{equation}
r_{MN}=\frac{C^\mathrm{E}_{MN}}{\sqrt{C^\mathrm{E}_{MM}C^\mathrm{E}_{NN}}}\;,
\end{equation} 
the behavior of the off-diagonal terms becomes clearer. (The
capital subscripts $N$ and $M$ can be different from the COSEBIs
subscripts, if several source populations are considered; see below
for more details.) \fig$\ref{RMT3theta}$ 
compares the correlation coefficients for three different choices of
the angular
range, $[1',400']$, $[20',400']$, and $[1',20']\thinspace$, at a
fixed $M=9$.

\begin{figure}
  \begin{center}
    \begin{tabular}{c}
      \resizebox{80mm}{!}{\includegraphics{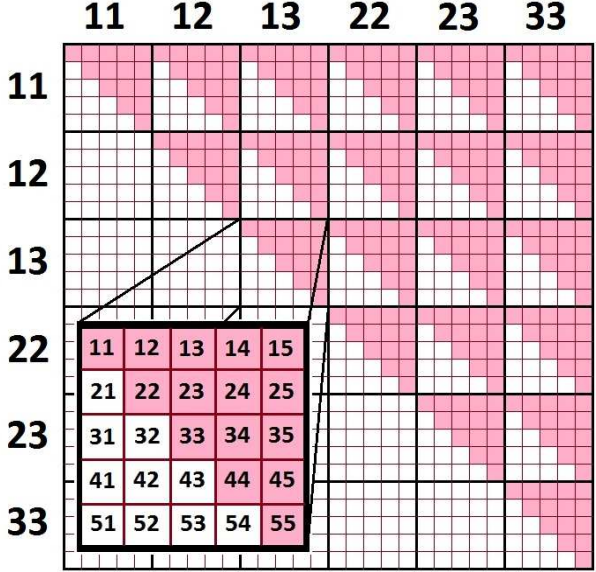}}\\
    \end{tabular}
    \caption{\small{A representation of a tomographic covariance. In this diagram 3 redshift-bins (1,2,3) and 5 COSEBIs modes are assumed to be present. The blow-up shows one of the covariance building blocks; the numbers 1-5 show the COSEBIs mode considered, e.g. 15 means the covariance of $E_1$ and $E_5$. The numbers on the sides of the matrix show which combination of redshift bins is considered, e.g., 12 means the covariance of redshift-bins 1 and 2 is relevant. Due to symmetry, only a part of the covariance elements have to be calculated, here shown in pink.}}
    \label{matrix}
  \end{center}
\end{figure}

\begin{figure}
  \begin{center}
    \begin{tabular}{c}
      \resizebox{85mm}{!}{\includegraphics{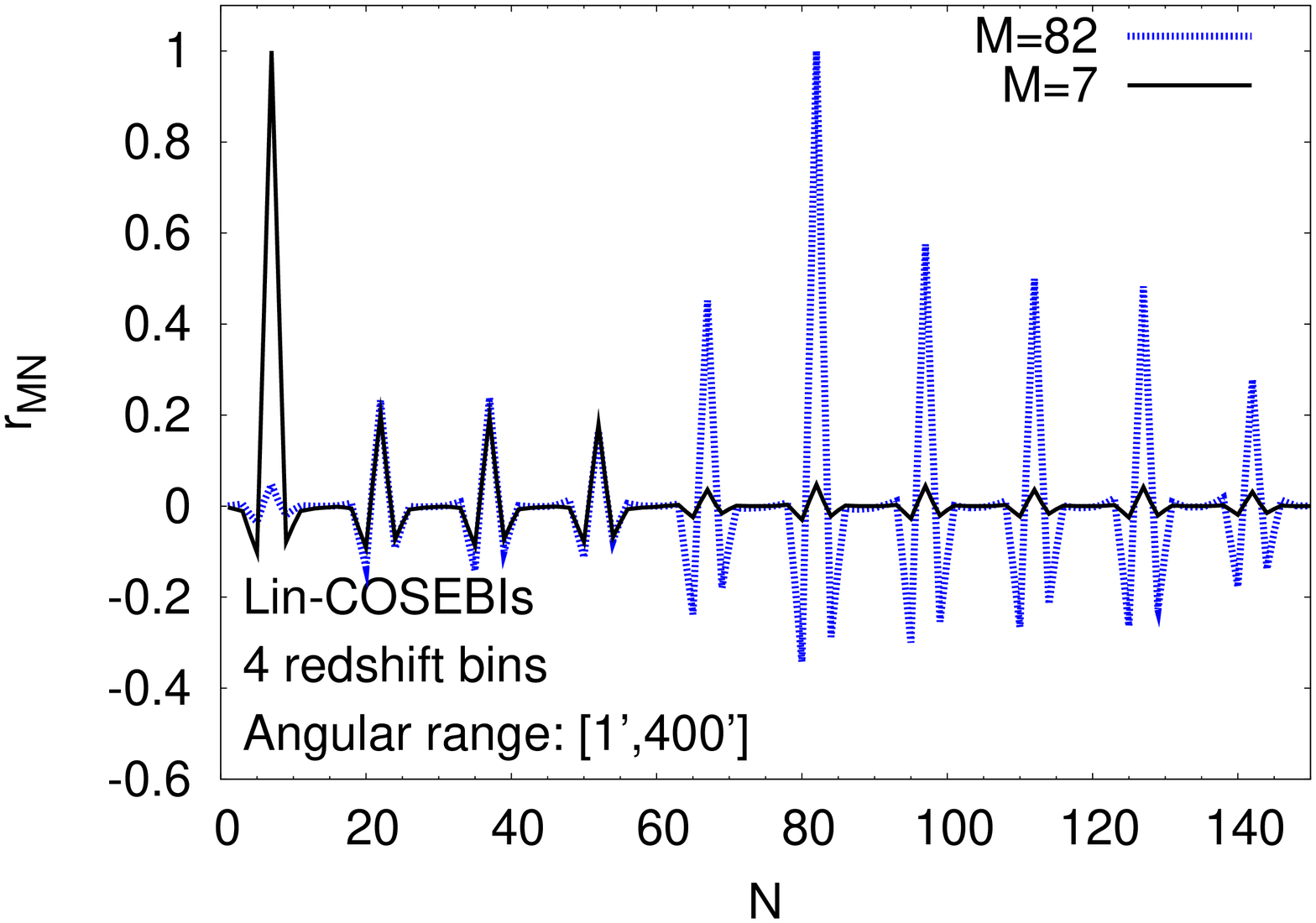}}\\
      \resizebox{85mm}{!}{\includegraphics{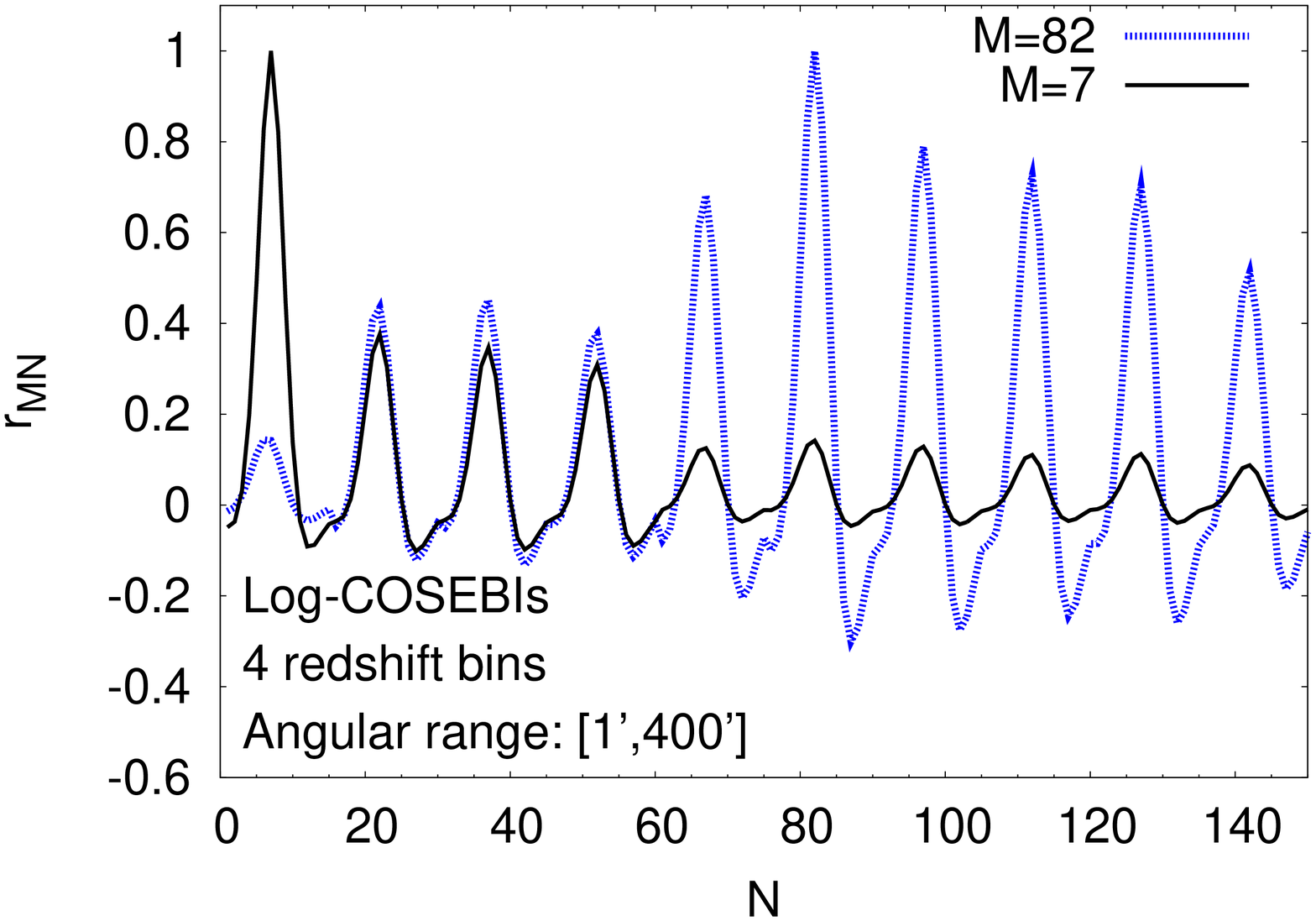}}
    \end{tabular}
    \caption{\small{The correlation coefficients of COSEBIs for an angular range of $[1',400']$ and 4 redshift bins. In total, 15 COSEBIs modes are considered for each graph. The $r_{MN}$ is shown for $M=7$ corresponding to $E_7^{11}$, and for $M=82$ corresponding to $E_7^{23}$.}}
    \label{RMTTomo}
  \end{center}
\end{figure}

\begin{figure}
  \begin{center}
    \begin{tabular}{c}
      \resizebox{85mm}{!}{\includegraphics{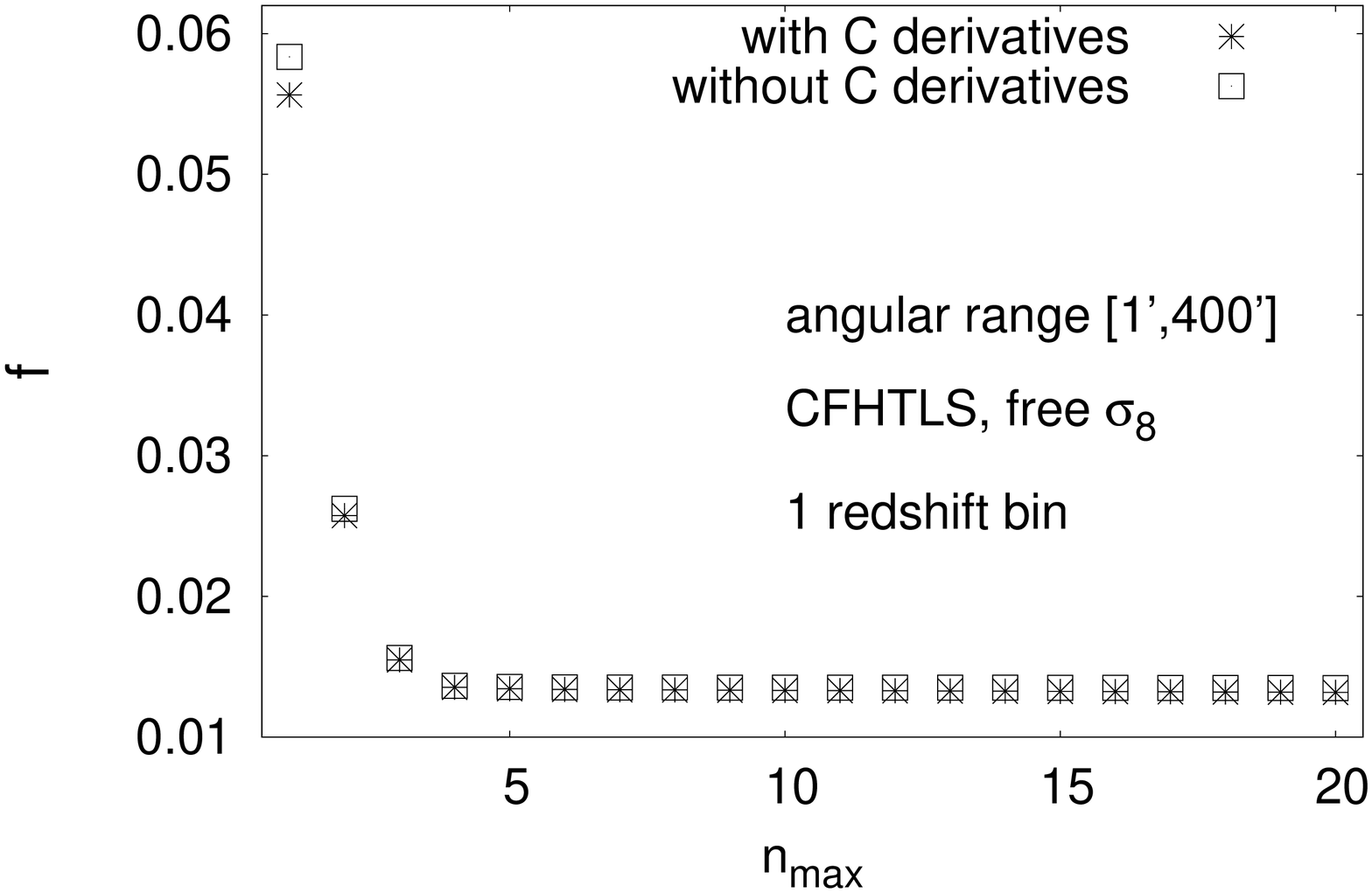}}\\
    \end{tabular}
    \caption{\small{A comparison between a simplified and complete
        Fisher analysis, using Log-COSEBIs. The asterisks show the
        case where the derivatives of the covariance is taken into
        account (the first part of \Eqt$\ref{FisherTeg}$) while the
        squares show the simplified case where we assume these
        derivatives are zero, in calculating $f$. Here $\sigma_8$ is
        the only free parameter, whereas the rest of the parameters
        is fixed to their fiducial values.}}
    \label{compare-F}
  \end{center}
\end{figure}

Cosmic shear analysis, as we will see in \sect$\ref{result}$, provides more
information when redshift information is available. In practice the
redshifts of galaxies are estimated using several photometric filters
(see e.g. \citealt{2010A&A...523A..31H}), from which an overall
distribution for the source galaxies is obtained. The distribution is
then divided into a number of photometric redshift bins. The photometric redshifts
are not exact, so the true redshift distributions will
overlap. Therefore, instead of redshift bins, in general one has to
consider redshift distributions. However, for simplicity, in the
present work we have assumed redshift bins with sharp cuts and no
overlap. In addition, the bins are selected such that the number of
galaxies in each bin is the same.

In general, a tomographic covariance for $r$ redshift bins consists of
$[r(r+1)/2]^2$ building blocks, each of which is a covariance
matrix of $(E^{ij}_n,E^{kl}_m)$ where $i,j,k,l$ are fixed and $n,m=1,2,...,n_\mathrm{max}$. This means in total the covariance matrix has
$[r(r+1)\:n_\mathrm{max}]^2/4$ elements, where $n_\mathrm{max}$ is the
maximum number of COSEBIs modes considered.

Nevertheless, a covariance
matrix is by definition symmetric and a tomographic covariance is made
up of smaller covariances, i.e., only $x(x+1)/2 \times
n_\mathrm{max}(n_\mathrm{max}+1)/2$ elements, with $x=r(r+1)/2$, have
to be calculated, the rest are equal to these (see
\fig$\ref{matrix}$).

The covariance of the $E^{ij}_n$ depends on six indices; in order to
apply normal matrix operations, the three indices of $E^{ij}_n$ are
combined into one `superindex' $N$, given by
\begin{equation}
 N=\Big[(i-1)\times r-\frac{(i-1)(i-2)}{2}+(j-i)\Big]\times n_\mathrm{max}+n\;,
\end{equation}
where $r$ is the total number of redshift bins and $n_\mathrm{max}$ is the total number of COSEBIs modes.

Using the new labeling, the correlation coefficients of $E^{11}_7$ and
$E^{23}_7$ (corresponding to $N=7$ and $N=82$, respectively) with the
other $E^{ij}_{n}$ is shown in \fig$\ref{RMTTomo}$, where 15 COSEBIs
modes and 4 redshift bins are considered.  Each of the peaks in the
figure correspond to the correlation coefficient of $E^{11}_7$ and
$E^{ij}_7$. The highest peak with $r=1$ occurs for $M=N$, while the rest
of the peaks are correlations between different redshift bins. The
Log-COSEBIs show larger noise-correlations between different modes compared
to Lin-COSEBIs, which may persuade one to choose the Lin-COSEBIs for
cosmic shear analysis. However, the Log-COSEBIs compensate this
apparent disadvantage by
requiring fewer modes to saturate the Fisher information level for relevant cosmological parameters
compared to the linear ones, i.e., the number of covariance elements
that have to be calculated for Lin-COSEBIs is higher and hence
determining their covariance matrix is more time consuming, especially
when redshift binning is considered. Consequently, in
\sect$\ref{result}$ we mainly employ Log-COSEBIs to analyze
tomographic Fisher information.

\section{Results of Fisher analysis}
\label{result}

\subsection{Figure-of-merit}

In this section we carry out a figure-of-merit analysis to demonstrate
the capability of COSEBIs to constrain cosmological parameters from
cosmic shear data. Our figure-of-merit, $f$, based on the Fisher
matrix, quantifies the credibility of the estimated parameters. In
general, for any unbiased estimator, the Fisher matrix gives the lower
limit of the errors on parameter estimations (see
e.g. \citealt{statistics2} and \citealt{stastistics} for details).

The Fisher matrix is related to the COSEBIs by
\begin{align}
\label{FisherTeg}
F_{ij}= \frac{1}{2} \mathrm{Tr}[C^{-1}\:C_{,i}\:C^{-1}\:C_{,j}+C^{-1}\:M_{ij}]\;,
\end{align} 
where $C$ is the COSEBIs covariance,
$M_{ij}=\boldsymbol{E}_{,i}\;\boldsymbol{E}_{,j}^\mathrm{T}+\boldsymbol{E}_{,j}\:\boldsymbol{E}_{,i}^\mathrm{T}$,
$\boldsymbol{E}$ is the vector of the E-mode COSEBIs, and the commas followed
by subscripts indicate partial derivatives with respect to the
cosmological parameters (see \citealt{1997ApJ...480...22T} for
example). We define our figure-of-merit, $f$, in a very similar manner
to SEK 
\begin{equation}
\label{f}
f=\left(\frac{1}{\sqrt{\mathrm{det}\: F}}\right)^{1/n_\mathrm{p}}\;,
\end{equation} 
where $n_\mathrm{p}$ is the number of free parameters considered. In
the following analysis, we will assume for simplicity that the first
term in \Eqt$\eqref{FisherTeg}$ is much smaller than the second and can
thus be neglected. Note that this approximation becomes more realistic
in the case of a large survey area, since the first term on the r.h.s.
of \Eqt$\eqref{FisherTeg}$ does not depend on the survey area, while
the second term is proportional to it (recall that $C
\varpropto 1/A$ or in other words $C^{-1} \varpropto A$). We checked that
our medium survey is already big
enough for this approximation to hold (see \fig$\ref{compare-F}$).

With the definition $\eqref{f}$ we compress the Fisher matrix into a
one-dimensional quantity, which provides a measure of the geometric
mean of the standard deviations of the parameters; e.g. in the case of
one free parameter $\phi$, $f(\phi)$ is equal to the standard
deviation $\sigma(\phi)$ of that parameter.\footnote{Another quantity, $q$,
was also defined in SEK to measure the area of the likelihood
regions. It is calculated from the second-order moments of the
posterior likelihood. $q$ and $f$ are equal if the posterior is a
multivariate Gaussian. 
\cite{2011MNRAS.418..536E} has shown that the difference
between $f$ and $q$ is small, especially for a large survey area.}

\begin{figure*}
\sidecaption
    \includegraphics[width=0.6\textwidth]{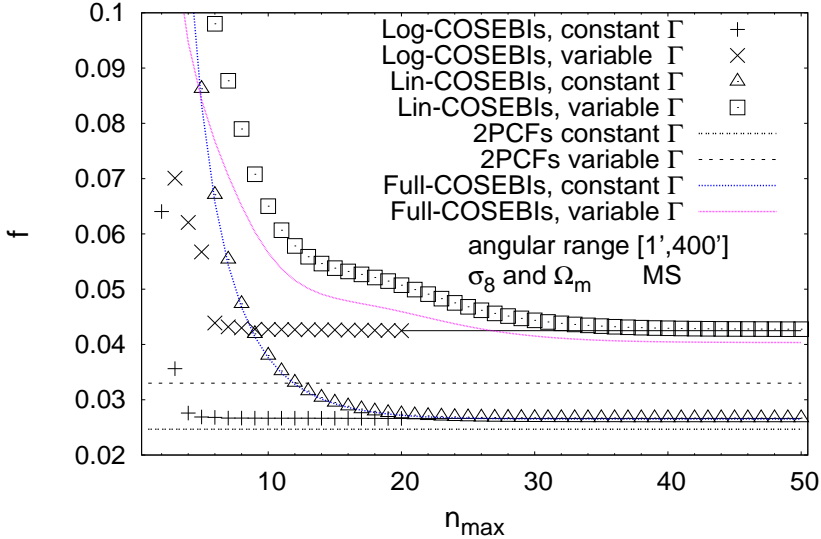}
    \caption{A comparison between the Log- and Lin-COSEBIs Fisher
      analysis results for two sets of assumptions, where $\sigma_8$
      and $\Omega_\mathrm{m}$ are the free parameters and the rest is
      fixed to their fiducial values. In one case the shape parameter
      $\Gamma$ is held fixed, while in the other it is left as a
      variable depending on $\Omega_\mathrm{m}$ and the fiducial
      values of $h$ and $\Omega_\mathrm{b}$ (according to \Eqt$\ref{shapeparameter}$). The
      same analysis is also carried out for the Full-COSEBIs and the shear
      2PCFs.}
    \label{comparison-2param}
\end{figure*}

\begin{figure*}
  \begin{center}
    \begin{tabular}{c}
      \resizebox{130mm}{!}{\includegraphics{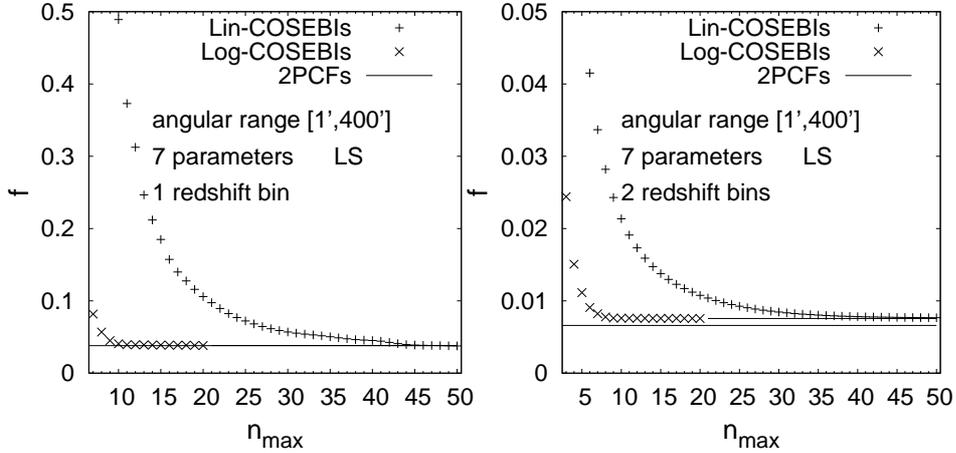}}
    \end{tabular}
    \caption{\small{These plots show one of our consistency checks. It
        is a comparison between the Lin- and Log-COSEBIs results for
        LS with a single (left panel) and two galaxy redshift
        distributions (right panel), including all of the 7
        parameters. Apart from very small numerical inaccuracies, both
        sets of COSEBIs saturate to the same value, as expected.
        There are two solid lines in each plot. The line with the
        higher value shows the value of Log-COSEBIs at
        $n_\mathrm{max}=20$, and the other line shows the value of $f$
        as obtained from the shear 2PCFs. The slightly smaller value
        of $f$ in the latter case (this difference is not visible in the plot) is related to the fact
        that the analysis from the shear 2PCFs implicitly assume the
        absence of B-modes, and thus contains information from very
        large-scale modes which, however, cannot be uniquely assigned
        to either E- or B-modes.  The comparison of the two plots
        shows that dividing the galaxies into two redshift bins not
        only increases the information content of the Fisher analysis
        but also decreases the number of COSEBIs modes needed. Note
        that the $x$-axis of the single redshift distribution plot
        starts from 7, the other one from 3.}}
    \label{comparison}
  \end{center}
\end{figure*}

\subsection{Assumptions and parameter settings}
\label{assump}

For our cosmic shear analysis we considered a medium (MS) and a large
survey (LS) as explained in \sect$\ref{CosmoModel}$. We have also
studied the effect of a Gaussian prior, in the form of a Fisher
matrix. This prior is the inverse of the WMAP7 parameter covariance
matrix from the final iteration (5000 sample points) of a Population
Monte Carlo (PMC) run (see \citealt{2010MNRAS.405.2381K}), called the
CMB prior from here on. We implement the prior by adding the Fisher
matrices of our COSEBIs analysis and the CMB prior. The value of the
CMB prior is shown in terms of $f(\phi)$ in the first column of
\tab$\ref{tablef}$ for each of the parameters.


We consider three different angular ranges, $[1',20']\:$,
$[1',400']\:$, and $[20',400']\:$. The motivation for this choice is
as follows: We consider a total interval of $[1',400']\:$ where the
flat sky approximation is still valid up to the maximum separation and
galaxy shapes are easily distinguishable for the minimum separation;
also, $\vartheta_{\rm min}=1'$ avoids the scales where baryonic
effects are expected to have the strongest ifluence. We further divide
this interval into two non-overlapping parts with $\vartheta_{\rm
  max}/\vartheta_{\rm min}=20$, to compare cosmic shear information on
small and large scales. The small-scale range, $[1',20']\:$ may apply
for a cosmic shear survey of individual one square degree fields. The
large scale interval, $[20',400']\:$ could be used for very
conservative analyses where non-linear and baryonic effects are to be
avoided.

In \sect$\ref{prop}$ we show the value of $f$ for two parameters while
the rest are fixed to their fiducial values for the MS, and also for
all seven parameters for the LS.  In principle we could show all of
the possible combinations for parameters, nevertheless finding the
error on each of the parameters seems a more relevant task. Therefore,
the rest of our analysis, carried out in \sect$\ref{forc}$, is done
for a single parameter, $\phi$, where $f(\phi)=\sigma_{\phi}$.

To find the value of $f$ for a single parameter we use two
approaches. In one approach we fix the six other parameters to their
fiducial values in \tab$\ref{cosmparam}$, while in the other case we
marginalized over the remaining six parameters.

For each setup we investigate the amount of information with respect
to the number of COSEBIs modes considered. In addition we analyze the
behavior of $f$ with the number of redshift bins considered.

\begin{figure*}
  \begin{center}
      \begin{tabular}{c}
         \resizebox{180mm}{!}{\includegraphics{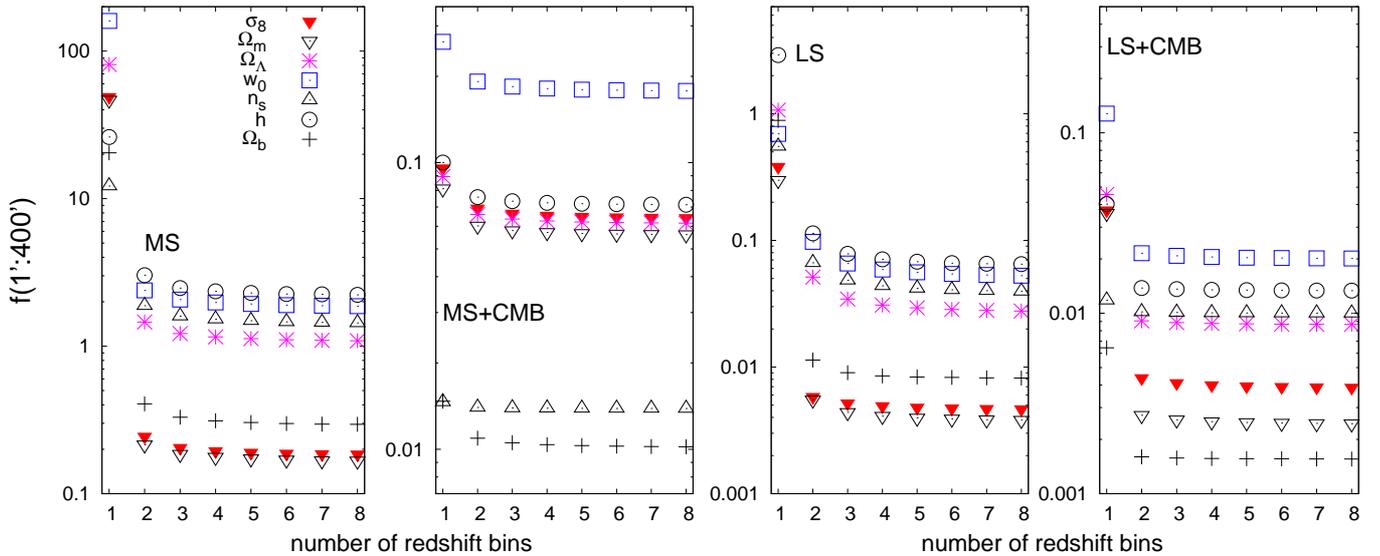}}\\
     \end{tabular}
     \caption{\small{The dependence of the COSEBIs saturated
         information content in form of $f$ for each of the seven
         parameters while the rest is marginalized over. The above
         plots are relevant to MS and LS parameters for an angular
         ranges of $[1',400']\:$. The redshift dependence of $f$ is
         very outstanding here and hence we needed to use logarithmic
         scales for the $y$-axes, especially in the cases where cosmic
         shear analysis is done without any priors.}}
   \label{fmAllMargin-redshift}
  \end{center}
\end{figure*}

\begin{figure*}
  \begin{center}
    \begin{tabular}{c}
      \resizebox{55mm}{!}{\includegraphics{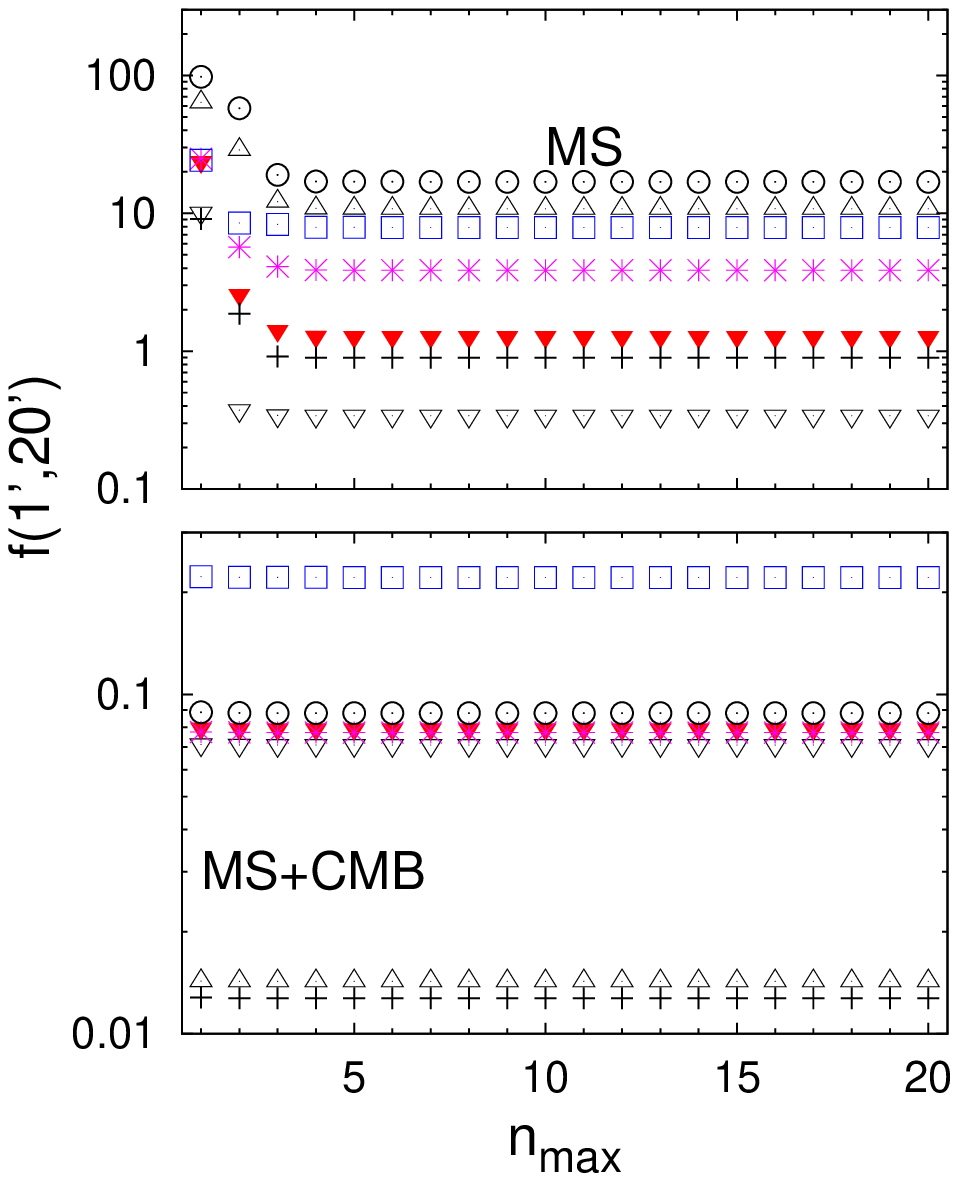}}
       \resizebox{55mm}{!}{\includegraphics{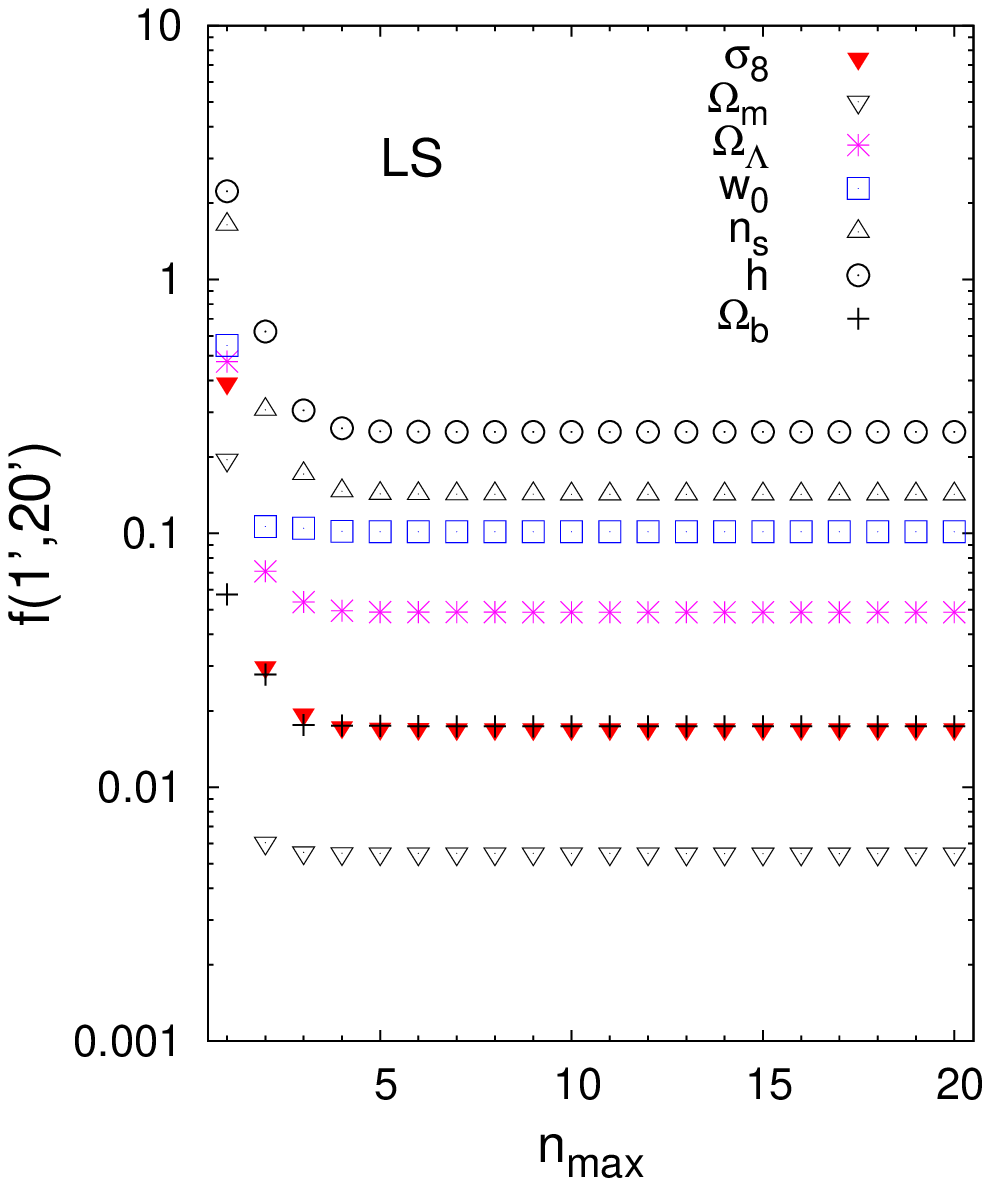}}
       \resizebox{55mm}{!}{\includegraphics{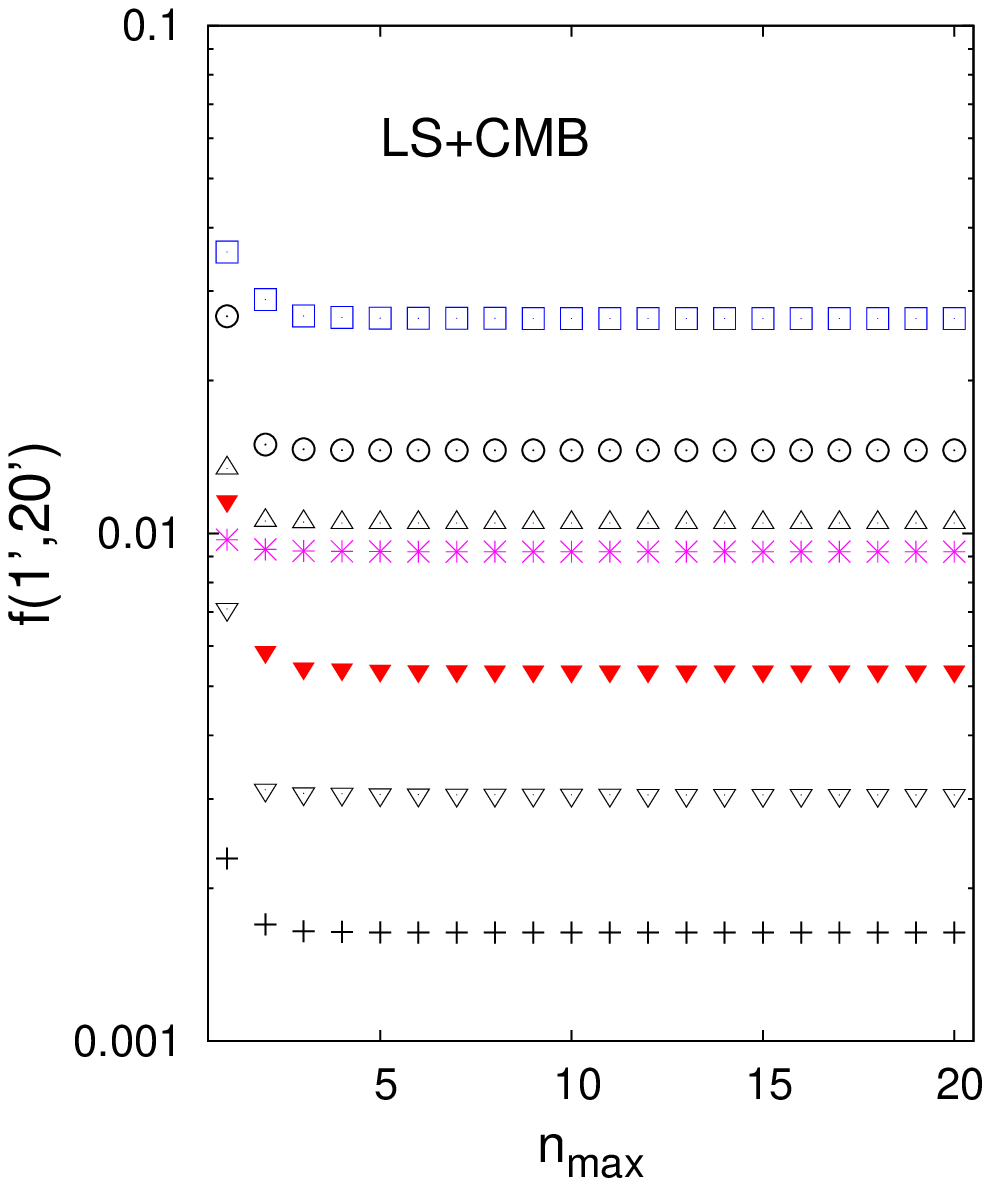}}\\
       \resizebox{55mm}{!}{\includegraphics{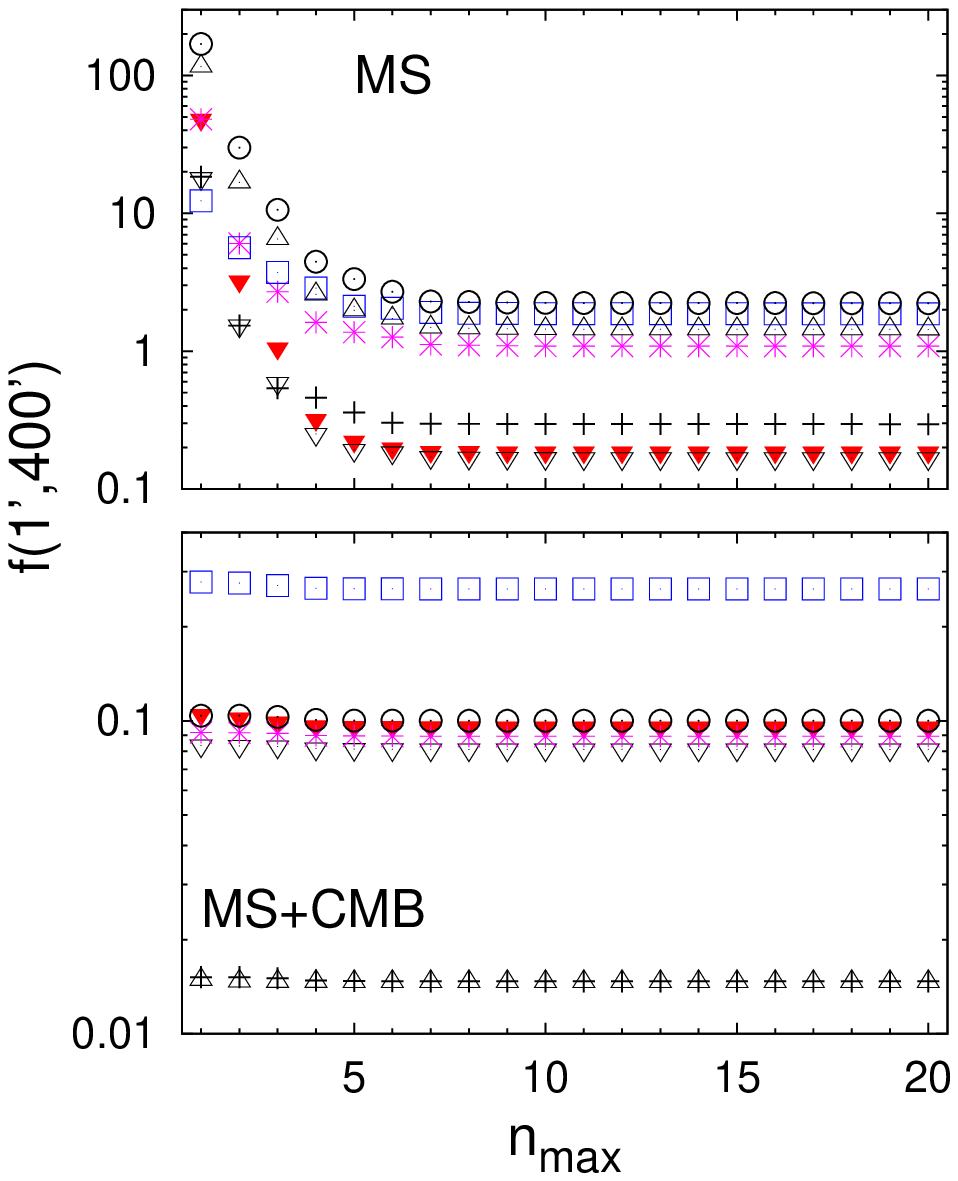}}       
       \resizebox{55mm}{!}{\includegraphics{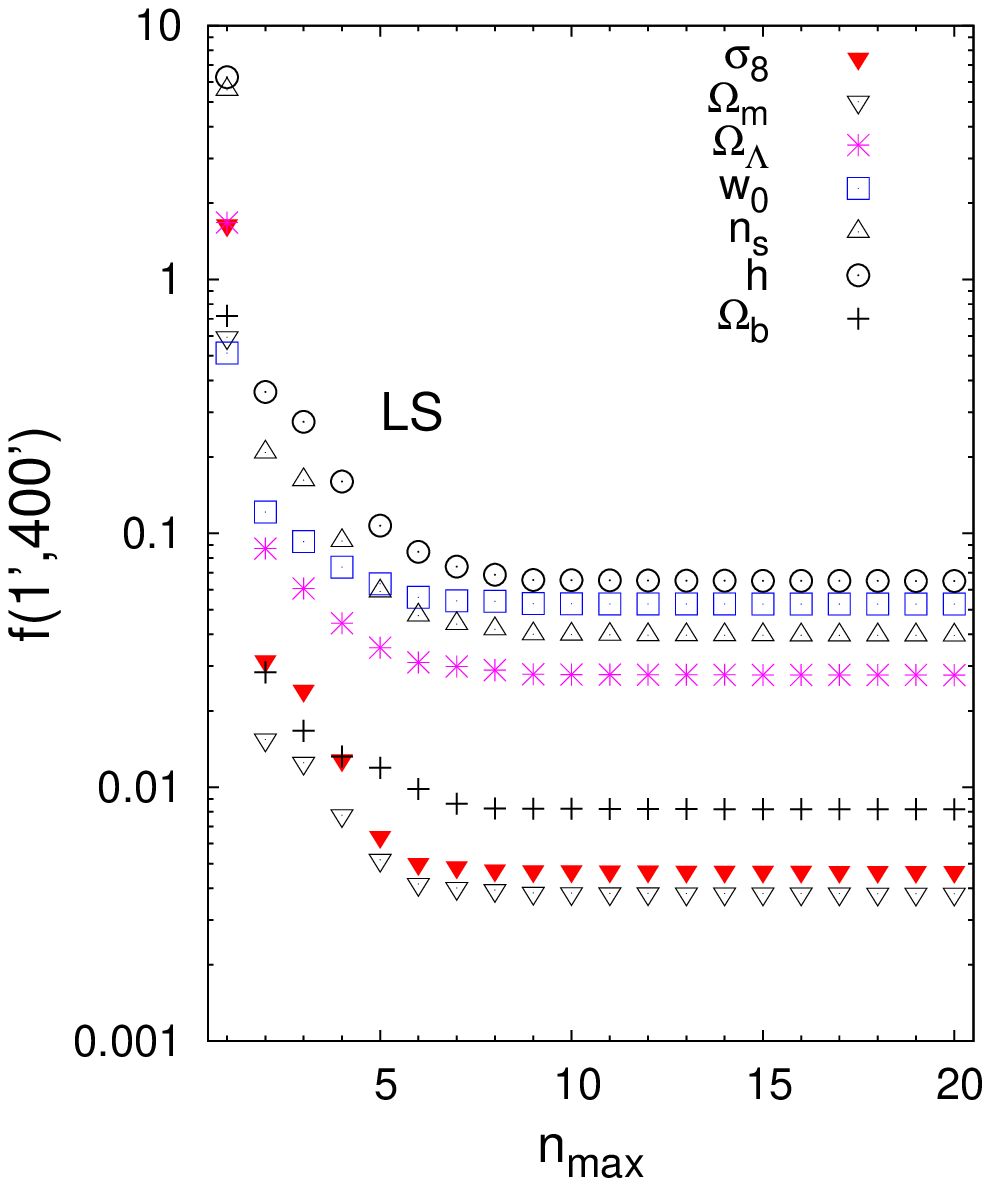}}
       \resizebox{55mm}{!}{\includegraphics{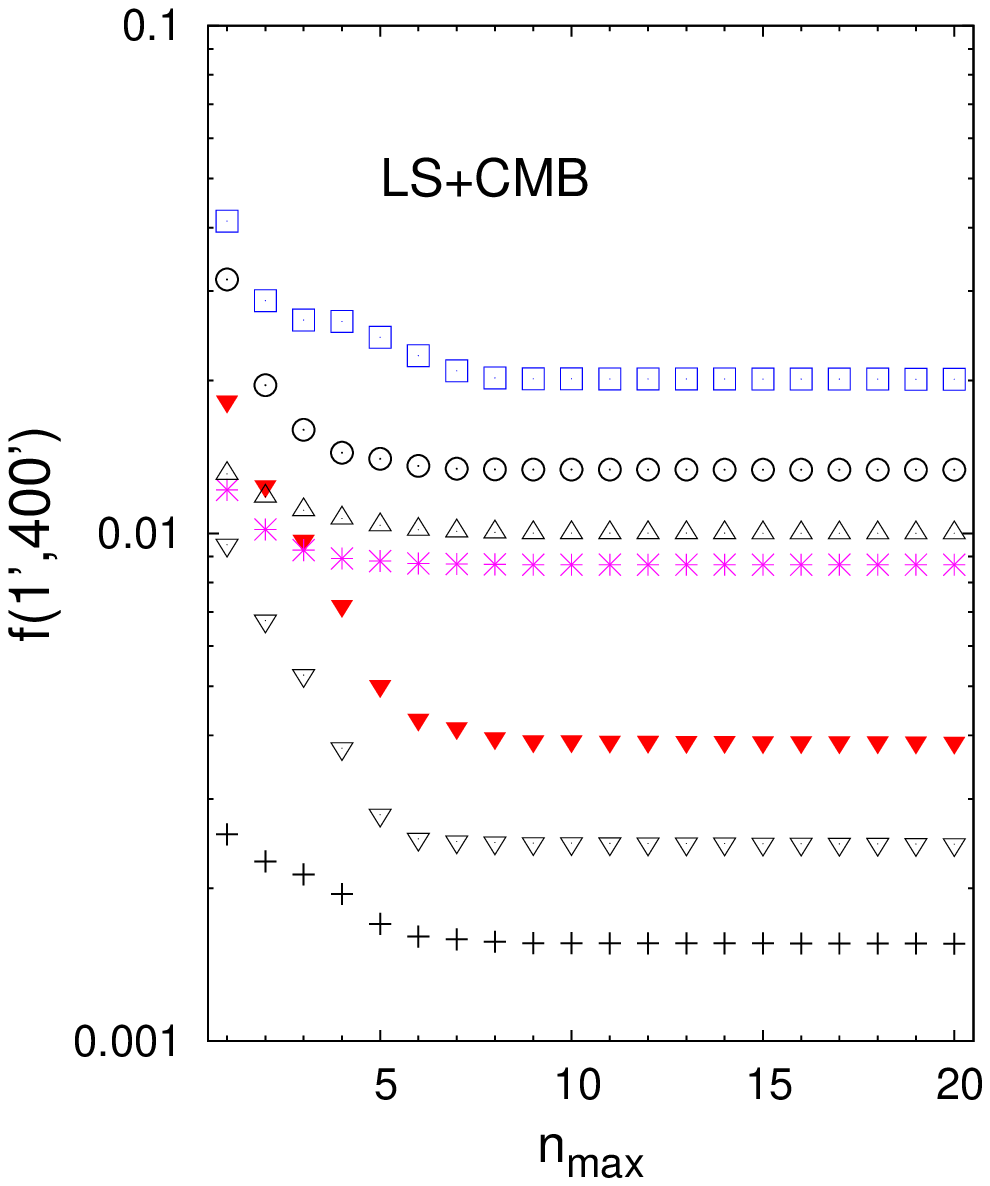}}\\
       \resizebox{55mm}{!}{\includegraphics{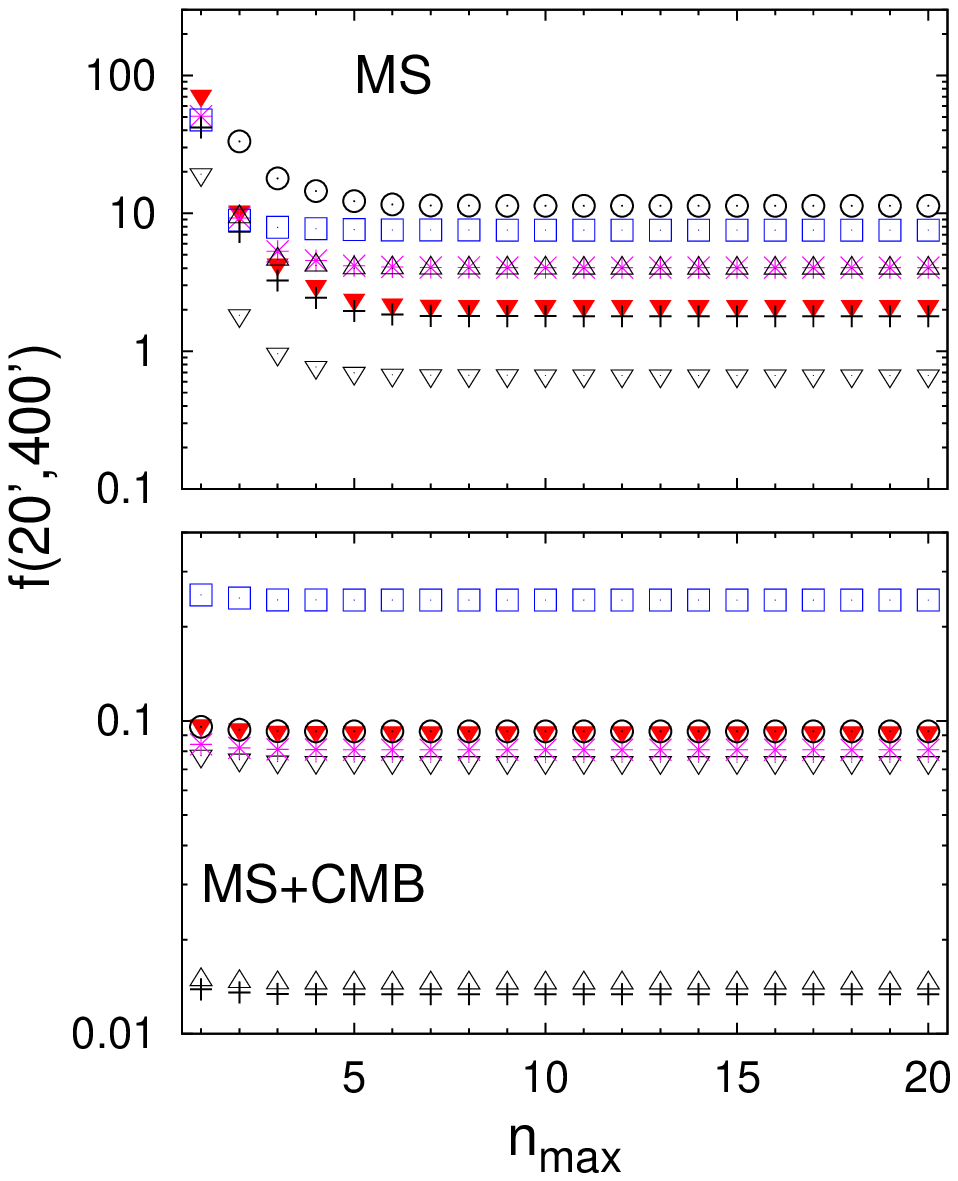}}       
      \resizebox{55mm}{!}{\includegraphics{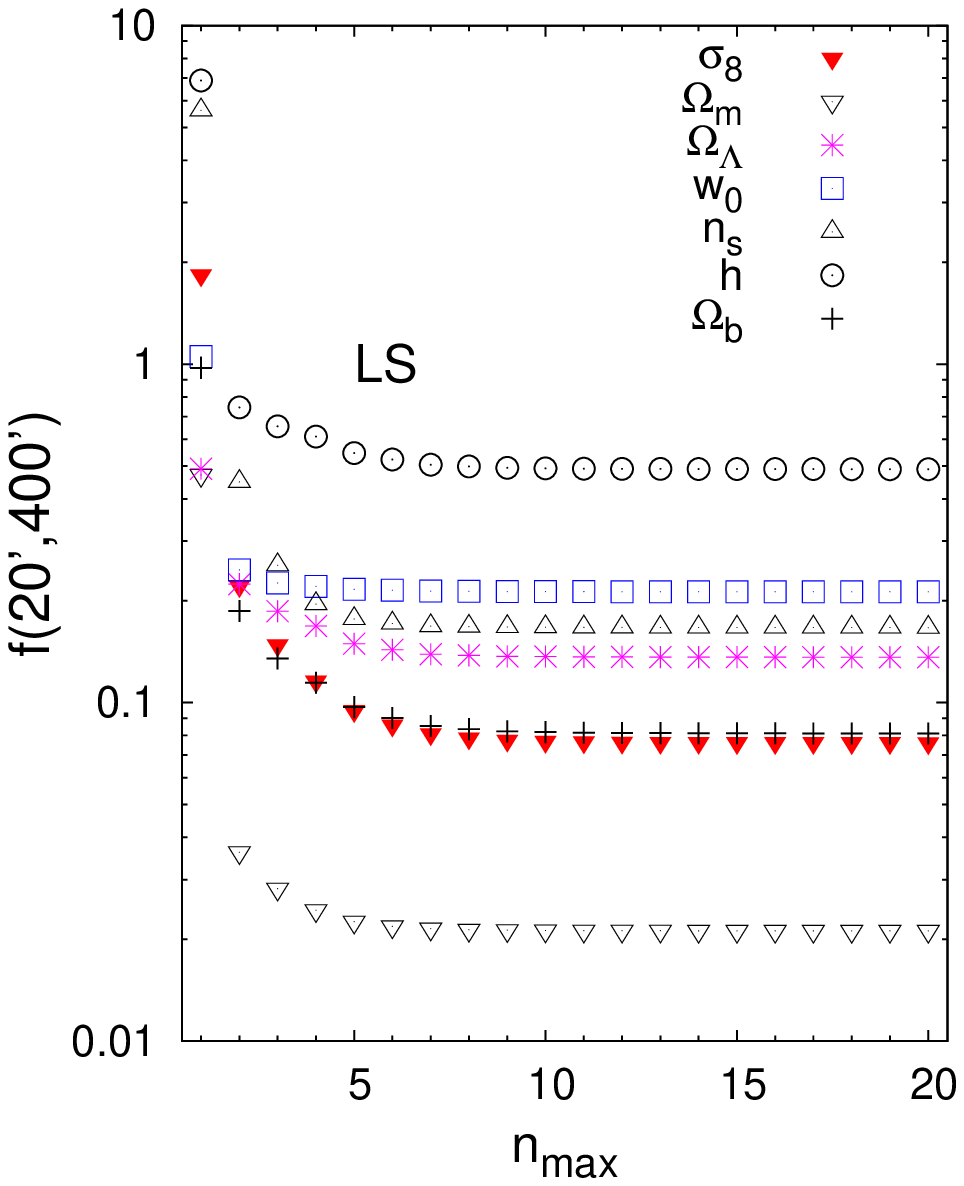}}
       \resizebox{55mm}{!}{\includegraphics{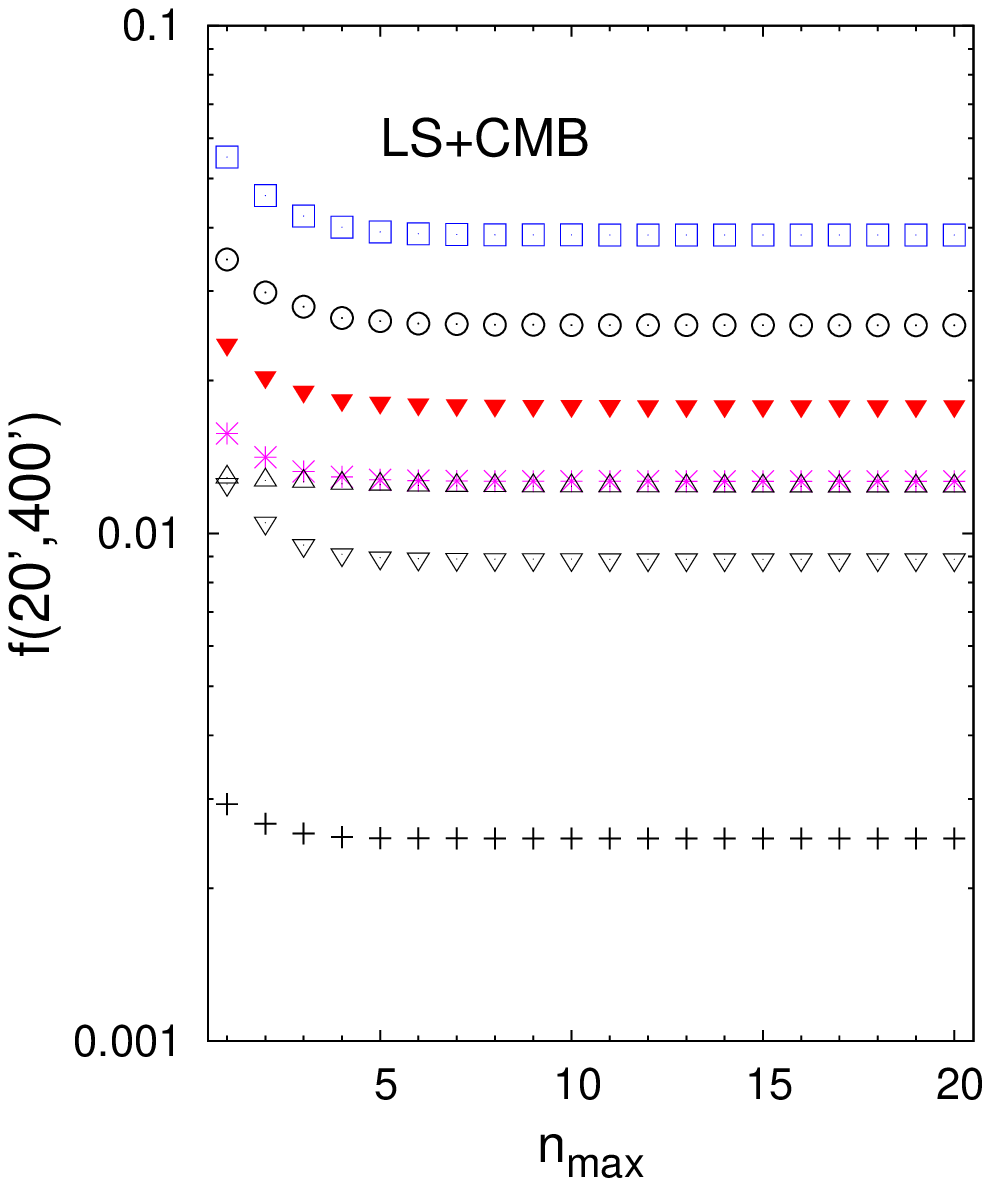}}\\
    \end{tabular}
    \caption{\small{Plots of figure-of-merit values with respect to the number of Log-COSEBIs modes considered for two surveys with the three angular ranges $[1',20']\:$, $[1',400']\:$, $[20',400']\:$. Here all the parameters except one is marginalized over. The first column correspond to MS (top plots) and MS+CMB (bottom plots) and the next two to LS and LS+CMB correspondingly. Eight redshift bins are used here. The CMB prior flattens the curves, especially in the case of MS, where WMAP7 puts tighter constrains on the parameters.}}
    \label{fm8bin}
  \end{center}
\end{figure*}

\subsection{Properties of COSEBIs}
\label{prop}
For a fixed number of modes, the Log-COSEBIs are more sensitive than
the Lin-COSEBIs to structures of the shear 2PCFs on small scales. Here
we show its effect on the Fisher analysis. We investigate the
dependence of $f$ on the number $n_{\rm max}$ of COSEBIs modes
incorporated in the analysis.

SEK have shown the difference between the behavior of the Lin- and
Log-COSEBIs for two parameters, $\sigma_8$ and $\Omega_\mathrm{m}$,
with $\Gamma$ fixed (their definition of $f$ and fiducial values of
parameters are slightly different from ours). Similar to their work,
we here compare the values of $f$ for the same two parameters with
Lin- and Log-COSEBIs. In addition, we inspect the difference between a
fixed shape parameter, or its dependence as given in \Eqt$\eqref{shapeparameter}$.

\fig$\ref{comparison-2param}$ is a representation of our inspection
for the MS in the angular range of $[1',400']\:$. Two general
conclusions come out of this comparison: (1) $f$ for fixed and
dependent $\Gamma$ converges to the same value for the Lin- and
Log-COSEBIs. (2) The values of $f$ for a fixed or dependent $\Gamma$
are different, and also the convergence rate is different.  E.g., the
Lin-COSEBIs reach the saturated $f$ value for $n_{\rm max} \approx 40$
for a variable $\Gamma$, while in the other case, only 25 modes are
needed. This effect is less dramatic in the case of Log-COSEBIs (they
need 7 modes for a variable $\Gamma$ and 5 modes for a constant one),
since they generally converge faster. Similarly in
\fig$\ref{comparison}$, we visualize our consistency check by showing
that the values of $f$ for Log- and Lin-COSEBIs converge to the same
value for seven parameters.\footnote{In general there are slight
  differences between the final value of $f$ due to numerical
  inaccuracies, but these differences never exceed a few percent and
  are typically much smaller. An exception happens when the saturation
  is too slow, and the Fisher matrix elements are too small, which is
  the case for MS with one redshift bin and 7 parameters, observable
  especially after marginalizing over 6 parameters when the remaining
  parameter is $w_0$ , $\Omega_\mathrm{m}$, $\Omega_\Lambda$ or
  $\sigma_8$. However, for these cases, $f$ is much larger than unity,
  i.e., cases in which no meaningful constraints can be obtained
  anyway.}

In Figs.\thinspace$\ref{comparison-2param}$ and $\ref{comparison}$, we
also show the value of $f$ as derived directly from the shear 2PCFs,
i.e., without \mbox{E-/B-mode} separation. As expected, in this case
$f$ becomes slightly smaller since it is now implicitly assumed that
all the signal is due to E-modes. However, this is not justified in
general; for example, very large-scale modes (i.e., small $\ell$)
enter $\xi_+(\vartheta)$ even for small $\vartheta$, and such modes
cannot be uniquely assigned to either E- or B-modes. Thus, the
decrease of $f$, and accordingly, the information gain is just an
apparent one, bought by making a strong assumption. The relative
difference between the 2PCFs and the converged Lin-/Log-COSEBIs values
for $f$ is larger for the variable $\Gamma$ case, since here
small-$\ell$ modes, which are filtered out in the COSEBIs, contain
information about the power spectrum shape.

We also considered as further possibility that the requirement of
finite support for the $\xi_-(\vartheta)$ is dropped, and call this
`Full-COSEBIs'. They form a complete set of functions on
$[\vartheta_{\rm min},\vartheta_{\rm max}]$, without the constraints
given in \Eqt$\eqref{T+cond}$.\footnote{They are obtained by adding
  two additional weight functions $T_+$ to those used in the COSEBIs;
  for the linear case, we just take all Legendre polynomials (see
  SEK).}  Though not physically reasonable, the Full-COSEBIs are
equivalent to measuring $\xi_+$ only, on the same interval. As can be
seen from \fig$\ref{comparison-2param}$, the full COSEBIs yield a
slightly lower value of $f$ than the true COSEBIs, showing that
$\xi_-$ on scales larger than $\vartheta_{\rm max}$ adds apparent
information, which, however, is not observable. We stress here that
the E-/B-mode correlation functions $\xi_{\rm E/B}$, introduced by
\cite{2002ApJ...568...20C} and \cite{2002A&A...396....1S}, are
essentially equivalent to the Full-COSEBIs, since they are also based
on the assumption that $\xi_-$ can be measured to arbitrarily large
separations -- which, however, is not possible. Therefore, a cosmic
shear analysis based on $\xi_{\rm E}$ (e.g.,
\citealt{2008A&A...479....9F}; \citealt{2011arXiv1111.6622L})
underestimates the uncertainties of cosmological parameters.

Furthermore, we compare the Lin- and Log-COSEBIs for LS parameters in
\fig$\ref{comparison}$, for one and two redshift bins. The $x$-axis in
the left plot starts from 7 in contrast to the right one which starts
from 3. The reason is that to constrain $n_\mathrm{p}$ parameters at
least $n_\mathrm{p}$ equations are needed, i.e., if one redshift bin
is considered, $n_\mathrm{p}$ COSEBIs modes should be accounted for to
produce a covariance matrix with at least $n_\mathrm{p}$ $\times$
$n_\mathrm{p}$ elements.  For more than one redshift bin, a smaller
number of COSEBIs modes are sufficient, subsequently the saturation
rate of $f$ is faster, as is visible in the right plot in the
figure. Recall that 2 redshift bins means 3 different redshift
combinations, i.e., for 7 parameters, the smallest integer not less
than $\lceil 7/3 \rceil = 3$ COSEBIs modes are needed.

\subsection{Forecast for parameter constraints}
\label{forc}

This section is dedicated to our final results according to the
assumptions and parameters explained in \sect$\ref{assump}$.

\fig$\ref{fmAllMargin-redshift}$ shows the dependence of $f$ for 20
Log-COSEBIs modes and for the $[1',400']$ angular range on the number
of galaxy distributions (i.e., redshift bins), where all but one
parameter are marginalized over. Dividing the galaxy distribution into
more than 4 redshift bins does not change the value of $f$
considerably. Nevertheless, a much larger number of redshift bins is
required to control and correct for systematic effects, e.g., coming
from intrinsic alignments (see for
example \citealt{2010arXiv1009.2024J} and references therein).

We also show the dependence of $f$ on $n_{\rm max}$, for 8 redshift
bins and marginalized parameters, in \fig$\ref{fm8bin}$. Comparing the
cosmic shear analysis with and without CMB prior, we see from the
figure that the prior in general flattens the curves. 
However, the
curves are flatter for MS+CMB than LS+CMB as a result of the larger
difference between the LS and the CMB prior.

The constraints on each of the cosmological parameters behave
differently with respect to the number of COSEBIs modes or redshift
bins considered. For marginalized parameters where the
behavior of parameters is entangled, their curves show a similar
decline.

By comparing the different angular ranges we conclude that a wider
angular range needs more modes to extract all information. We also
note that the behavior of the seven parameters are not similar and
each of them should be followed separately.

Based on the results from these two figures, we will report additional
results for $n_{\rm max}=20$, where the value of $f$ is converged, and
for either one or eight redshift bins.  These results are shown in
\tab$\ref{tablef}$ in the form of $f(\phi)$ for different cases. We
have compared these values with \cite{2010MNRAS.404..110D}, and found
them fully consistent.

In the following we explain our conclusions from the two mentioned
figures and \tab$\ref{tablef}$ in more detail:

\begin{table*}
  \caption{Listed below are the values of $f$ for 20 Log-COSEBIs
    modes, where the saturation level is well reached for all the
    cases. The first column indicates which parameter is free, the
    rest of the parameters are either fixed (4th-7th columns) or
    marginalized over (8th-11th columns). The two numbers written in
    the first-column box of each parameter in parenthesis are the
    corresponding $f$ 
    values from CMB alone, for fixed and marginalized parameters,
    respectively. The second and third columns show the survey and
    angular range considered. The values of $f$ for one and eight
    redshift bins are presented here for comparison. MS and LS stand
    for a medium and large survey, respectively.} 
\begin{center}
\footnotesize{\begin{tabular}{ccc|c|c||c|c||c|c||c|c|}
\cline{4-11}
& & & \multicolumn{4}{|c||}{Fixed parameters} & \multicolumn{4}{|c|}{Marginalized parameters}\\
\cline{4-11}
& & & \multicolumn{2}{|c|}{without Prior} 
& \multicolumn{2}{|c||}{+ CMB}
& \multicolumn{2}{|c|}{without Prior} 
& \multicolumn{2}{|c|}{+ CMB}
\\ \cline{4-11}
& & & 1 z-bin & 8 z-bins &  1 z-bin & 8 z-bins & 1 z-bin & 8 z-bins &  1 z-bin & 8 z-bins \\ \cline{1-11}
\multicolumn{1}{|c|}{\multirow{3}{*}{ \large$\Omega_\mathrm{b}$}} &
\multicolumn{1}{|c|}{\multirow{3}{*}{MS}} &
         {$[1',20']$}      &    3.61E$-$2	&	3.30E$-$2	&	9.39E$-$4	&	9.39E$-$4	&  2.79E$+$2	&	8.95E$-$1	&	1.52E$-$2	&	1.27E$-$2
 \\ \cline{3-11}
\multicolumn{1}{|c|}{\multirow{5}{*} { (9.39E$-$4) }} &
\multicolumn{1}{|c|}{}      &
        { $[1',400']$}      &    3.28E$-$2	&	2.83E$-$2	&	9.39E$-$4	&	9.39E$-$4	&  2.04E$+$1	&	2.95E$-$1	&	1.47E$-$2	&	1.02E$-$2
 \\ \cline{3-11}
\multicolumn{1}{|c|}{}      &
\multicolumn{1}{|c|}{}      &
        { $[20',400']$}      &    1.07E$-$1	&	6.88E$-$2	&	9.39E$-$4	&	9.39E$-$4	&  1.26E$+$2 &	1.79E$+$0	&	1.51E$-$2	&	1.34E$-$2
 \\ \cline{2-11}
\multicolumn{1}{|c|}{\multirow{3}{*} {(1.52E$-$2)}} &
\multicolumn{1}{|c|}{\multirow{3}{*}{LS}} &
        { $[1',20']$}  &    7.39E$-$4	&	6.80E$-$4	&	5.81E$-$4	&	5.51E$-$4	&   4.65E+0	&	1.74E$-$2	&	8.67E$-$3	&	1.64E$-$3
  \\ \cline{3-11}
\multicolumn{1}{|c|}{}      &
\multicolumn{1}{|c|}{}      &
         $[1',400']$  &   7.15E$-$4	&	6.41E$-$4	&	5.69E$-$4	&	5.29E$-$4	&  8.86E$-$1	&	8.19E$-$3	&	6.44E$-$3	&	1.56E$-$3
  \\ \cline{3-11}
\multicolumn{1}{|c|}{}      &
\multicolumn{1}{|c|}{}      &
         $[20',400']$  &    7.07E$-$3	&	3.31E$-$3	&	9.31E$-$4	&	9.04E$-$4	&  	5.01E+0 &	8.11E$-$2 &	1.35E$-$2	&	2.51E$-$3
 \\ \hline \hline
\multicolumn{1}{|c|}{\multirow{3}{*}{\large$h$ }} &
\multicolumn{1}{|c|}{\multirow{3}{*}{MS}} &
         $[1',20']$      &   1.32E$-$1	&	1.19E$-$1	&	2.09E$-$2	&	2.08E$-$2	&  3.38E+2	&	1.69E+1	&	1.04E$-$1	&	8.81E$-$2
\\ \cline{3-11}
\multicolumn{1}{|c|}{\multirow{5}{*} { (2.11E$-$2) }} &
\multicolumn{1}{|c|}{}      &
         $[1',400']$      &   1.21E$-$1	&	1.06E$-$1	&	2.08E$-$2	&	2.07E$-$2	&  2.61E+1	&	2.23E+0	&	1.00E$-$1	&	7.13E$-$2
\\ \cline{3-11}
\multicolumn{1}{|c|}{}      &
\multicolumn{1}{|c|}{}      &
         $[20',400']$      &   3.91E$-$1	&	2.78E$-$1	&	2.11E$-$2	&	2.11E$-$2	& 1.11E+3	&	1.13E+1	&	1.03E$-$1	&	9.26E$-$2
\\ \cline{2-11}
\multicolumn{1}{|c|}{\multirow{3}{*} {(1.04E$-$1)}} &
\multicolumn{1}{|c|}{\multirow{3}{*}{LS}} &
         $[1',20']$      &   2.73E$-$3	&	2.52E$-$3	&	2.71E$-$3	&	2.50E$-$3	&   1.70E+1	&	2.51E$-$1	&	5.18E$-$2	&	1.46E$-$2
 \\ \cline{3-11}
\multicolumn{1}{|c|}{}      &
\multicolumn{1}{|c|}{}      &
         $[1',400']$      &    2.65E$-$3	&	2.40E$-$3	&	2.63E$-$3	&	2.38E$-$3	 &   2.91E+0	&	6.49E$-$2	&	4.02E$-$2	&	1.33E$-$2
 \\ \cline{3-11}
\multicolumn{1}{|c|}{}      &
\multicolumn{1}{|c|}{}      &
         $[20',400']$      &   2.67E$-$2	&	1.37E$-$2	&	1.66E$-$2	&	1.15E$-$2	&  1.81E+1	&	4.89E$-$1 &	9.80E$-$2	&	2.57E$-$2
 \\ \hline \hline
\multicolumn{1}{|c|}{\multirow{3}{*}{\large$n_\mathrm{s}$}} &
\multicolumn{1}{|c|}{\multirow{3}{*}{MS}} &
         $[1',20']$      & 1.07E$-$1	&	9.80E$-$2	&	7.81E$-$3	&	7.81E$-$3	& 7.53E+2	&	1.08E+1	&	1.47E$-$2	&	1.43E$-$2
\\ \cline{3-11}
\multicolumn{1}{|c|}{\multirow{5}{*}{(7.83E$-$3)}} &
\multicolumn{1}{|c|}{}      &
         $[1',400']$      & 9.08E$-$2	&	7.87E$-$2	&	7.80E$-$3	&	7.79E$-$3	&  1.22E$+$1	&	1.44E$+$0	&	1.46E$-$2	&	1.39E$-$2
 \\ \cline{3-11}
\multicolumn{1}{|c|}{}      &
\multicolumn{1}{|c|}{}      &
         $[20',400']$      &  1.95E$-$1	&	1.54E$-$1	&	7.83E$-$3	&	7.82E$-$3	& 1.04E$+$2	&	3.97E$+$0	&	1.50E$-$2	&	1.45E$-$2
 \\ \cline{2-11}
\multicolumn{1}{|c|}{\multirow{3}{*} {(1.58E$-$2)}} &
\multicolumn{1}{|c|}{\multirow{3}{*}{LS}} &
         $[1',20']$      &  2.12E$-$3	&	1.98E$-$3	&	2.05E$-$3	&	1.92E$-$3	& 8.47E$+$0	&	1.42E$-$1	&	1.28E$-$2	&	1.05E$-$2
\\ \cline{3-11}
\multicolumn{1}{|c|}{}      &
\multicolumn{1}{|c|}{}      &
         $[1',400']$      &  2.03E$-$3	&	1.83E$-$3	&	1.96E$-$3	&	1.79E$-$3	& 5.52E$-$1	&	3.96E$-$2	&	1.18E$-$2	&	1.00E$-$2
\\ \cline{3-11}
\multicolumn{1}{|c|}{}      &
\multicolumn{1}{|c|}{}      &
         $[20',400']$      &  1.29E$-$2	&	8.15E$-$3	&	6.70E$-$3	&	5.65E$-$3	& 4.11E$+$0 &	1.67E$-$1 &	1.30E$-$2	&	1.24E$-$2
\\ \hline \hline
\multicolumn{1}{|c|}{\multirow{3}{*}{\large$w_0$ }} &
\multicolumn{1}{|c|}{\multirow{3}{*}{MS}} &
         $[1',20']$      &  1.40E$-$1	&	1.18E$-$1	&	3.64E$-$2	&	3.59E$-$2	&  2.70E$+$2	&	7.89E$+$0	&	2.70E$-$1	&	2.21E$-$1
 \\ \cline{3-11}
\multicolumn{1}{|c|}{\multirow{5}{*}{(3.77E$-$2)}} &
\multicolumn{1}{|c|}{}      &
         $[1',400']$      & 1.02E$-$1	&	8.69E$-$2	&	3.54E$-$2	&	3.46E$-$2	& 1.60E$+$2	&	1.87E$+$0	&	2.64E$-$1	&	1.78E$-$1
 \\ \cline{3-11}
\multicolumn{1}{|c|}{}      &
\multicolumn{1}{|c|}{}      &
         $[20',400']$      &  3.55E$-$1	&	3.06E$-$1	&	3.75E$-$2	&	3.74E$-$2	 &  4.11E$+$2	&	7.55E$+$0	&	2.80E$-$1	&	2.43E$-$1
 \\ \cline{2-11}
\multicolumn{1}{|c|}{\multirow{3}{*} {(2.83E$-$1)}} &
\multicolumn{1}{|c|}{\multirow{3}{*}{LS}} &
         $[1',20']$      &   3.04E$-$3	&	2.72E$-$3	&	3.03E$-$3	&	2.71E$-$3	&  3.59E+1	&	1.01E$-$1	&	1.68E$-$1	&	2.65E$-$2
 \\ \cline{3-11}
\multicolumn{1}{|c|}{}      &
\multicolumn{1}{|c|}{}      &
         $[1',400']$     &  2.83E$-$3	&	2.49E$-$3	&	2.82E$-$3	&	2.48E$-$3	&  6.93E$-$1	&	5.26E$-$2	&	1.28E$-$1	&	2.01E$-$2
 \\ \cline{3-11}
\multicolumn{1}{|c|}{}      &
\multicolumn{1}{|c|}{}      &
         $[20',400']$      &  2.12E$-$2	&	1.73E$-$2	&	1.85E$-$2	&	1.57E$-$2	& 3.75E+1 &	 2.12E$-$1 &	2.44E$-$1	&	3.87E$-$2
 \\ \hline \hline
\multicolumn{1}{|c|}{\multirow{3}{*}{\large$\Omega_\Lambda$}} &
\multicolumn{1}{|c|}{\multirow{3}{*}{MS}} &
         $[1',20']$      & 1.73E$-$1	&	1.26E$-$1	&	1.18E$-$2	&	1.18E$-$2	& 3.04E+2	&	3.86E+0	&	9.23E$-$2	&	7.72E$-$2
 \\ \cline{3-11}
\multicolumn{1}{|c|}{\multirow{5}{*}{(1.18E$-$2)}} &
\multicolumn{1}{|c|}{}      &
         $[1',400']$     &  9.56E$-$2	&	7.37E$-$2	&	1.17E$-$2	&	1.17E$-$2	& 8.11E+1	&	1.09E+0	&	8.94E$-$2	&	6.15E$-$2
\\ \cline{3-11}
\multicolumn{1}{|c|}{}      &
\multicolumn{1}{|c|}{}      &
         $[20',400']$      &  2.41E$-$1	&	1.98E$-$1	&	1.18E$-$2	&	1.18E$-$2	& 1.38E+3	&	4.04E+0	&	9.12E$-$2	&	8.09E$-$2
\\ \cline{2-11}
\multicolumn{1}{|c|}{\multirow{3}{*} {(9.26E$-$2)}} &
\multicolumn{1}{|c|}{\multirow{3}{*}{LS}} &
         $[1',20']$      &  3.47E$-$3	&	2.70E$-$3	&	3.33E$-$3	&	2.63E$-$3	& 1.50E+1	&	4.89E$-$2	&	5.86E$-$2	&	9.20E$-$3
\\ \cline{3-11}
\multicolumn{1}{|c|}{}      &
\multicolumn{1}{|c|}{}      &
         $[1',400']$     & 2.99E$-$3	&	2.28E$-$3	&	2.90E$-$3	&	2.23E$-$3	& 1.07E+0	&	2.77E$-$2	&	4.55E$-$2	&	8.67E$-$3
 \\ \cline{3-11}
\multicolumn{1}{|c|}{}      &
\multicolumn{1}{|c|}{}      &
         $[20',400']$      & 1.34E$-$2	&	1.02E$-$2	&	8.85E$-$3	&	7.70E$-$3	&   1.86E+1 &	1.36E$-$1	&	8.18E$-$2	&	1.27E$-$2
 \\ \hline \hline
\multicolumn{1}{|c|}{\multirow{3}{*}{\large$\Omega_\mathrm{m}$}} &
\multicolumn{1}{|c|}{\multirow{3}{*}{MS}} &
         $[1',20']$      & 9.32E$-$3	&	8.26E$-$3	&	4.06E$-$3	&	3.96E$-$3	& 9.27E+1	&	3.40E$-$1	&	8.44E$-$2	&	7.09E$-$2
 \\ \cline{3-11}
\multicolumn{1}{|c|}{\multirow{5}{*}{(4.52E$-$3)}} &
\multicolumn{1}{|c|}{}      &
         $[1',400']$     & 7.13E$-$3	&	6.47E$-$3	&	3.81E$-$3	&	3.70E$-$3	&  4.67E+1	&	1.67E$-$1	&	8.11E$-$2	&	5.61E$-$2
 \\ \cline{3-11}
\multicolumn{1}{|c|}{}      &
\multicolumn{1}{|c|}{}      &
         $[20',400']$      & 3.45E$-$2	&	3.23E$-$2	&	4.48E$-$3	&	4.47E$-$3	& 3.40E+2	&	6.67E$-$1	&	8.28E$-$2	&	7.39E$-$2
 \\ \cline{2-11}
\multicolumn{1}{|c|}{\multirow{3}{*} {(8.56E$-$2)}} &
\multicolumn{1}{|c|}{\multirow{3}{*}{LS}} &
         $[1',20']$      &  2.18E$-$4	&	2.03E$-$4	&	2.18E$-$4	&	2.02E$-$4	&  9.15E+0	&	5.48E$-$3	&	4.83E$-$2	&	3.05E$-$3
 \\ \cline{3-11}
\multicolumn{1}{|c|}{}      &
\multicolumn{1}{|c|}{}      &
         $[1',400']$     &  2.06E$-$4	&	1.90E$-$4	&	2.06E$-$4	&	1.90E$-$4	&  2.98E$-$1	&	3.81E$-$3	&	3.58E$-$2	&	2.44E$-$3
 \\ \cline{3-11}
\multicolumn{1}{|c|}{}      &
\multicolumn{1}{|c|}{}      &
         $[20',400']$      &  2.20E$-$3	&	1.96E$-$3	&	1.98E$-$3	&	1.79E$-$3	 & 3.81E+0 &	2.12E$-$2	&	7.50E$-$2	&	8.89E$-$3
 \\ \hline \hline
\multicolumn{1}{|c|}{\multirow{3}{*}{\large$\sigma_8$}} &
\multicolumn{1}{|c|}{\multirow{3}{*}{MS}} &
         $[1',20']$      & 1.75E$-$2	&	1.51E$-$2	&	1.12E$-$2	&	1.05E$-$2	& 1.01E+2	&	1.27E+0	&	9.79E$-$2	&	7.91E$-$2
 \\ \cline{3-11}
\multicolumn{1}{|c|}{\multirow{5}{*}{(1.46E$-$2)}} &
\multicolumn{1}{|c|}{}      &
         $[1',400']$     & 1.35E$-$2	&	1.20E$-$2	&	9.90E$-$3	&	9.27E$-$3	& 4.90E+1	&	1.85E$-$1	&	9.57E$-$2	&	6.42E$-$2
 \\ \cline{3-11}
\multicolumn{1}{|c|}{}      &
\multicolumn{1}{|c|}{}      &
         $[20',400']$      &  6.97E$-$2	&	6.33E$-$2	&	1.43E$-$2	&	1.42E$-$2	& 3.41E+2	&	2.14E+0	&	1.07E$-$1	&	9.23E$-$2
\\ \cline{2-11}
\multicolumn{1}{|c|}{\multirow{3}{*} {(1.11E$-$1)}} &
\multicolumn{1}{|c|}{\multirow{3}{*}{LS}} &
         $[1',20']$      &  4.05E$-$4	&	3.65E$-$4	&	4.04E$-$4	&	3.65E$-$4	& 1.01E+1	&	1.70E$-$2	&	5.12E$-$2	&	5.35E$-$3
\\ \cline{3-11}
\multicolumn{1}{|c|}{}      &
\multicolumn{1}{|c|}{}      &
         $[1',400']$      & 3.82E$-$4	&	3.43E$-$4	&	3.82E$-$4	&	3.43E$-$4	& 3.82E$-$1	&	4.66E$-$3	&	3.73E$-$2	&	3.88E$-$3
 \\ \cline{3-11}
\multicolumn{1}{|c|}{}      &
\multicolumn{1}{|c|}{}      &
         $[20',400']$      &  4.34E$-$3	&	3.58E$-$3	&	4.16E$-$3	&	3.48E$-$3	 &  4.50E+0 &	7.61E$-$2	&	8.99E$-$2	&	1.78E$-$2
\\ \cline{1-11}
\end{tabular}}
\end{center}
\label{tablef}
\end{table*}

\begin{itemize}
  
\item MS vs. LS vs. CMB prior: In general, because of its much larger
  survey area and larger galaxy number density, LS puts tighter
  constraints on all of the parameters than the MS. Furthermore, since
  the LS is deeper than the MS, it allows more sensitive constraints on
  parameters which are sensitive to the growth of structure, in
  particular $w_0$. As can be seen from \fig$\ref{fm8bin}$, the requested number
  of COSEBIs for saturation is slightly higher for the LS since this survey
  contains more information, but smaller than 20 in all cases.

  The MS constraints on parameters are weaker than the CMB alone for
  almost all cases. $\Omega_\mathrm{m}$ and $\sigma_8$, the two
  parameters for which present cosmic shear studies provide the
  most relevant constraints, are the only two for which the MS constraints are
  comparable with CMB. As for the rest of the parameters, except for
  $f(w_0,[1',400'])$ for the case of fixed parameters, the parameter
  uncertainty of the CMB prior is about one order of magnitude or more
  smaller than that of the MS. We conclude that the MS is not large
  enough to be competitive with the CMB for constraining a seven
  parameter cosmological model. However, the combination of the two
  slightly tightens the constraints.

  In contrast to the MS, the LS yields parameter constraints which can
  be considerably stronger than the CMB alone, in particular when
  tomography is employed. To wit, for the total angular
  range of $[1',400']$ and 8 $z$-bins, the LS errors are smaller than
  those from the CMB, except for $n_\mathrm{s}$ with marginalized
  parameters. For fixed parameters, the $f$ value resulting from the
  combination of the LS and the CMB prior is very close to the LS
  value, whereas it can be much smaller than that of the CMB
  alone. I.e., we conclude that the resulting constraints from the LS
  are not dominated by the assumed prior.

  In contrast to the MS, the LS is able to put useful constraints on
  the parameters, even without tomography and for marginalized
  parameters. This is seen in the difference between the error values
  obtained from CMB alone and LS+CMB.

  The relative value of errors on parameters is different between the
  two surveys. The reason is that the redshift distributions of the
  two surveys are different (see \tab$\ref{surveyTable}$ and
 \fig$\ref{pofz-Euclid-CFHTLS}$) as seen in the figures. Using
  redshift information in general is equivalent to using structure
  evolution information. The large-scale evolution is more visible in
  the case of a wider redshift distribution which starts from $z=0$,
  where these structures are more evolved. 

\item Fixed vs. Marginalized: The difference between the value of $f$
  for fixed and marginalized parameters is immense, especially when
  prior information is not available.  
  However, we can state that, for all cases without priors,
  $\Omega_\mathrm{m}$ is the best constrained parameter. The relative
  value of the parameters is also different between fixed and
  marginalized cases. Also 
  for the fixed parameter case, the convergence rate of LS is slower
  compared to the MS and its relative information content is higher, as
  expected.

  In contrast to the fixed parameters case, tomography substantially
  lowers the errors for marginalized parameters.

\item Angular ranges: From the table, we find the following common
  trends for fixed parameters:
\begin{align}
& f(1',400') < f(1',20') < f(20',400')\;,\\
& \left(\frac{f(1',20')}{f(1',400')}\right)_\mathrm{MS} > \left(\frac{f(1',20')}{f(1',400')}\right)_\mathrm{LS}\;,\\
& \left(\frac{f(20',400')}{f(1',400')}\right)_\mathrm{MS} < \left(\frac{f(20',400')}{f(1',400')}\right)_\mathrm{LS}\;.
\end{align} 
The first relation shows that there is more information at smaller
scales than in the large-scale angular interval, although there is
some independent information at larger scales. More interesting are
the second and third relations which show that there is more
information at smaller scales regarding LS compared to MS, which is a
consequence of their different redshift distributions and cuts (see
\fig$\ref{pofz-Euclid-CFHTLS}$ and \tab$\ref{surveyTable}$). The
relations between the $f(\vartheta_\mathrm{min},
\vartheta_\mathrm{max})$ for different angular ranges change when
parameters are marginalized over. In this case the above inequalities
are no longer valid for all of the parameters.

\end{itemize}

\begin{figure}
  \begin{center} 
  \begin{tabular}{l}
        \resizebox{83mm}{!}{\includegraphics{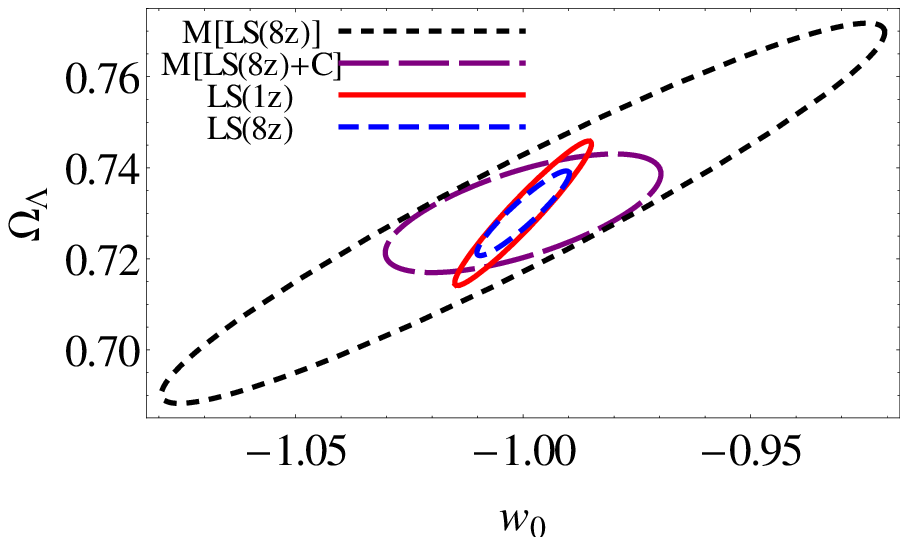}} \\
         \resizebox{86mm}{!}{\includegraphics{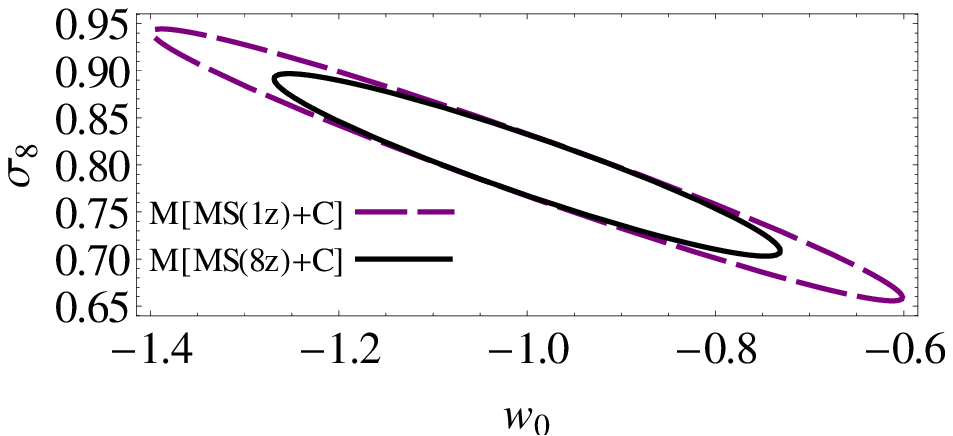}} 
   \end{tabular}
   \caption{\small{Contour plots of 1$\sigma$ constraints for pairs of
       parameters. The plot legends are abbreviated to save space. M
       stands for Marginalized, C for the CMB prior, and \#z for the
       number of redshift bins. Top plot: The 1$\sigma$ contours of
       $w_0$ and $\Omega_\Lambda$ for the LS. The two smaller contours
       are for the case where the five other parameters are held fixed
       to their fiducial values. Bottom plot: The 1$\sigma$ contours
       of $\sigma_8$ and $w_0$ for the MS.}}
    \label{conto}
  \end{center}
\end{figure}

\fig$\ref{conto}$ shows constraints for two pairs of parameters. In the
bottom plot the direction of the contours for the MS are determined by
the CMB prior. In this case tomography slightly improves the
constraints on both parameters. The top plot, on the other hand, shows
the constraints on dark energy parameters. In this case tomographic
improvements are more visible. Also here the direction of the contours
are different for the two cases.


%

\section{Summary and Conclusion}
\label{discussion}

We have generalized the cosmic shear analysis with COSEBIs to include seven
cosmological parameters, and investigated the effect of tomography on
parameter constraints. For our analysis we mainly used the
Log-COSEBIs, although for consistency checks we have shown that
Lin-COSEBIs results agree with their logarithmic
counterparts. In \App\ref{linearCOSEBIs} we show that there are
analytic solutions for linear $W_n$ functions, which are in principle
less computationally demanding. 

Besides the number of parameters and the use of redshift information,
the main difference to SEK is a technical one: we calculated the
COSEBIs and their covariance directly from the power spectrum, without
using the 2PCFs as intermediate step. This choice is more convenient
for theoretical considerations and highly speeds up the calculations
of the COSEBIs covariances, an advantage in particular for the case of
tomography.  We also confirmed that our method can reproduce the
results of SEK.

We investigated the effect of a Gaussian prior on two mock surveys,
using three angular ranges on which the shear 2PCFs are asumed to be
measured, by Fisher analysis methods. We considered the case that all
but one parameter are fixed, as well as that where we marginalize
over the other six parameters, in order to find the constraints on a
single parameter.  The prior was the Fisher matrix resulting from a
Population Monte Carlo (PMC) analysis of WMAP7 results. We considered
a medium and a large mock survey resembling the state-of-the-art in
cosmic shear and that of future all-(extragalactic) sky surveys. 

Most importantly, we found that a relatively small number of COSEBIs
captures essentially all cosmological information from a cosmic shear
survey. Whereas this number is larger than in SEK, due to the
higher-dimensional parameter space and the tomographic analysis,
COSEBIs not only act as a clean E-/B-mode separating shear statistics,
but also as a highly efficient data compression method. We stress that
this feature is extremely useful also for evaluating covariances from
numerical simulations.

The required number of COSEBIs to saturate the cosmological
information is considerably smaller for Log-COSEBIs than for
Lin-COSEBIs, which implies a clear preference for the former.  It also
increases with the number of free parameters, and depends on the
parameters considered, the survey, and the angular range on which the
2PCFs are measured. In all cases we considered, fewer than 20
Log-COSEBIs modes were sufficient to reach information saturation.
Aside from the tighter constraints of the large survey (LS) on all of the
parameters compared to the medium-sized survey (MS), the order of the
parameters with respect to their Fisher information is different
between the two surveys. Moreover, in general LS requires more COSEBIs
modes due to its higher information level. The comparison of the three
angular ranges shows that most of the information in cosmic shear is
contained at smaller scales which is why the Log-COSEBIs with their
finer oscillations towards smaller scales are more sensitive and reach
the saturated level of information with fewer modes. However, there is
interesting independent information at larger scales, resulting in 
tighter constraints when using both angular ranges. 

We have investigated the dependence of our figure of merit, $f$, on the
number of redshift bins considered. In agreement with earlier work, we
found that tomography greatly tightens parameter constraints, but the
cosmological information saturates at around three or four redshift
bins. This, however, does not imply that coarse redshift information
is sufficient for future lensing surveys, since good redshift
information is required to eliminate systematics from the data, such
as intrinsic alignment effects (e.g., \citealt{2003A&A...398...23K}
; \citealt{2008A&A...488..829J}; \citealt{2010A&A...523A...1J})

For future work it will be interesting to investigate the effects of
nulling with COSEBIs. Nulling techniques (\citealt{2008A&A...488..829J})
eliminate the intrinsic-intrinsic and intrinsic-shear correlations
from observed ellipticity correlations. The intrinsic-intrinsic
correlation can be handled by accurate redshift information to
eliminate pairs with physical connections. Consequently, one can
investigate how the cosmic shear information evolves by doing so, and
how many redshift bins are needed in this case.
Furthermore, our assumption of a Gaussian covariance becomes
unrealistic at small angular scales; hence, it will be interesting to
carry out a similar analysis based on more realistic covariances,
either obtained from ray-tracing through cosmological density fields
or using (semi-)analytic models, such as based on log-normal fields
(see \citealt{2011A&A...536A..85H}).


\begin{acknowledgements}
  We thank Benjamin Joachimi and Tim Eifler for interesting discussions. 
  We also like to thank Martin Kilbinger for sending us his CMB parameter covariance. 
  This work was supported by the Deutsche Forschungsgemeinschaft within the
  Transregional Research Center TR33 `The Dark Universe' and the
  Priority Programme 1177 `Galaxy Evolution' under the project SCHN
  342/9.
\end{acknowledgements}

\appendix
\section{Analytic solutions to linear COSEBIs weight functions}
\label{linearCOSEBIs}

Calculating the $W_n$ functions and evaluating integrals involving them requires careful methods, as a result of their very oscillating nature. In this section we first show our semi-analytical solutions to $W_n^\mathrm{Lin}(\ell)$ and at the end discuss our method of integration in more detail.

The filters, $T_{+n}^\mathrm{Lin}$, were calculated in SEK. By a simple variable change of $y=\ell\vartheta$, \Eqt$\eqref{Wn}$ becomes 
\begin{equation}
W_n^\mathrm{Lin}(\ell)=\frac{1}{\ell^2}\int_{\ell\vartheta_{\mathrm{min}}}^{\ell\vartheta_{\mathrm{max}}}\mathrm{d}y\:y\: T_{+n}(y/\ell) \mathrm{J}_0(y)\;.
\end{equation} One can write $T_{+n}$ in the form

\begin{equation}
\label{Tcoefs}
T_{+n}(\vartheta)=\sum_{i=0}^{n+1}a_{ni}\vartheta^i\;,
\end{equation} and then rewrite $W_n^\mathrm{Lin}$ as:

\begin{equation}
\label{WnFromSn}
W_n^\mathrm{Lin}(\ell)=\sum_{i=0}^{n+1}\frac{a_{ni}}{\ell^{2+i}}\int_{\ell\vartheta_{\mathrm{min}}}^{\ell\vartheta_{\mathrm{max}}}\mathrm{d}y\: y^{i+1} \mathrm{J}_0(x)\;.
\end{equation} We define the functions $S_n$ as

\begin{equation}
S_n(b):=\int_{0}^{b}\mathrm{d}x\: y^{n+1} \mathrm{J}_0(y)\;.
\end{equation} Inserting the above equation into \Eqt$\eqref{WnFromSn}$ gives

\begin{equation}
W_n^\mathrm{Lin}(\ell)=\sum_{i=0}^{n+1}\frac{a_{ni}}{\ell^{2+i}}[S_i(\ell\vartheta_{\mathrm{max}})-S_i(\ell\vartheta_{\mathrm{min}})]\;.
\end{equation} The $S_n$ functions can be obtained using standard Bessel functions relations, which in our case specialize to 
\begin{align}
\label{J1}
\frac{\mathrm{d}}{\mathrm{d}y}[y\: \mathrm{J}_1(y)]=y\:\mathrm{J}_{0}(y)\;,\\
\mathrm{J}_{-1}(y)-\mathrm{J}_{1}(y)=2\mathrm{J}'_0(y)\;,\nonumber\\
\mathrm{J}_{-1}(y)=-\mathrm{J}_1(y)\;,\\
\label{J2}
-\mathrm{J}_1(y)=\mathrm{J}'_0(y)\;,
\end{align} where the last equation results from the two equations before it.

\begin{figure}
  \begin{center}
      \resizebox{90mm}{!}{\includegraphics{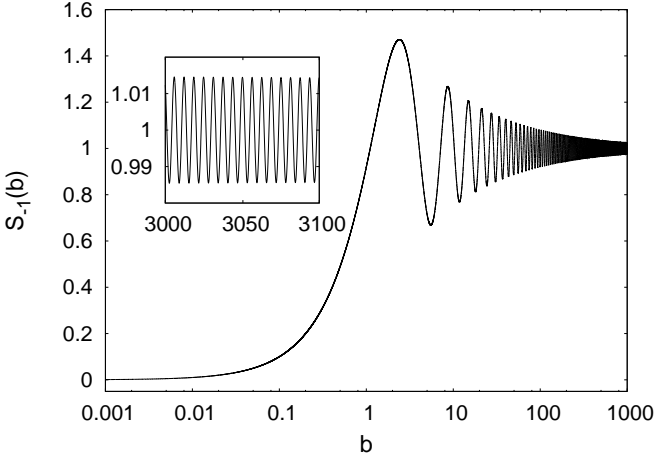}} 
    \caption{\small{The weight functions $W_n(\ell)$ depend on $S_n$ integrals, which are recursively related to each other. Here the shape of the starting function $S_{-1}$ is shown. As is expected it reaches zero as the argument gets smaller and has a highly oscillating behavior similar to its integrand, $\mathrm{J}_0$.}}
    \label{S-1}
  \end{center}
\end{figure}

Using Eqs.\thinspace$\eqref{J1}$ and $\eqref{J2}$, the following calculations for $S_n$ are carried out,

\begin{align}
S_n(b) & =\int_{0}^{b}\mathrm{d}y\: y^{n+1} \mathrm{J}_0(y) \nonumber\\
       & =  y^{n+1}\mathrm{J}_1(y)\Bigr\rvert_0^b-n\int_{0}^{b}\mathrm{d}x\:y^n\:\mathrm{J}_1(y)\nonumber\\
       & =  y^{n+1}\mathrm{J}_1(y)\Bigr\rvert_0^b+n\:y^n\mathrm{J}_0(y)\Bigr\rvert_0^b-n^2\int_{0}^{b}\mathrm{d}y\: y^{n-1} \mathrm{J}_0(y)\;,
\end{align} where we have carried out integration by parts twice. Note that the last term in the above equations is equal to $S_{n-2}$. Consequently, the recursive formula for $S_n$ is

\begin{equation}
\label{Snrec}
S_n(b)=b^{n+1}\mathrm{J}_1(b)+n\:b^n\mathrm{J}_0(b)-n^2\:S_{n-2}(b)\;.
\end{equation} The first two of these functions are

\begin{align}
S_0(b) & =\int_{0}^{b}\mathrm{d}y\: y \:\mathrm{J}_0(y)= b\:\mathrm{J}_1(b)\;,\\
S_{-1}(b) & =\int_{0}^{b}\mathrm{d}y\:\mathrm{J}_0(y)=b\; _1F_2[{1/2},{1,3/2},-b^2/4]\;,
\end{align} where $_1F_2$ is the Hypergeometric functions (see
\citealt{arfken}). Although there is an analytic formula for $S_{-1}$,
it is more convenient to solve it numerically using a similar
step-wise integration method as was explained in \sect$\ref{filters}$
. The only difference here is that
the steps are taken between zeros of the Bessel function. A plot of
$S_{-1}$ can be seen in \fig$\ref{S-1}$.

The recursive formula in \Eqt$\eqref{Snrec}$ can be rewritten as a closed form formula of the form:

\begin{align}
\label{Snclosed}
S_n(b) & = \mathrm{J}_1(b)[b^{n+1}+\sum_{k=1}^{N}(-1)^k\prod_{i=0}^{k-1}(n-2i)^2 b^{n+1-2k}] \nonumber\\
       & +\mathrm{J}_0(b)[nb^n+\sum_{k=1}^{N}(-1)^k\prod_{i=0}^{k-1}(n-2i)^2(n-2k) b^{n-2k}]\nonumber\\
       & +(-1)^{N+1}\prod_{i=0}^{N}(n-2i)^2 S_{n-2(N+1)}(b)\;,
\end{align} where $N$ is 

\begin{equation}
N  =
\left\{
\begin{aligned}
  & \frac{n-2}{2}\qquad n \;\; \mathrm{even}\\
  & \frac{n-1}{2}\qquad n \;\; \mathrm{odd} \;.
\end{aligned}
\right.
\end{equation}

To calculate the $T_{+n}^\mathrm{Lin}(\vartheta)$ SEK rescaled the angular separation intervals from $[\vartheta_\mathrm{min}, \vartheta_\mathrm{max}]$ to $[-1,1]$ with a variable transformation of the form

\begin{equation}
\label{xtheta}
 x=\frac{2(\vartheta-\bar{\vartheta})}{\Delta\vartheta}\;,
\end{equation} with $\bar{\vartheta}=(\vartheta_{\mathrm{min}}+\vartheta_{\mathrm{max}})/2$, $\Delta\vartheta=\vartheta_{\mathrm{min}}-\vartheta_{\mathrm{max}}$ and calculated the results for $t_{+n}^\mathrm{Lin} (x)=T_{+n}^\mathrm{Lin}(\vartheta)$. The explicit mathematical form of $t_{+1}^\mathrm{Lin}(x)$ and $t_{+2}^\mathrm{Lin}(x)$ are shown in SEK. One can find the $a_{ni}$ coefficients in \Eqt$\eqref{Tcoefs}$ by a change of variable from $x$ to $\vartheta$.

The rest of the $t_{+n}^\mathrm{Lin}(x)$ filters are

\begin{equation}
 t_{+n}^\mathrm{Lin}(x)=\sqrt{\frac{2n+3}{2}}\; P_{n+1}(x)\;,
\end{equation} where $P_{n+1}(x)$ is the Legendre polynomials of order $n+1$. We rewrite the recursive relation for Legendre polynomials using \Eqt$\eqref{xtheta}$

\begin{align}
P_0(x)  &=1\;,\nonumber \\
P_1(x)  &=\frac{2\vartheta}{\Delta \vartheta}-\frac{1}{B}\;, \nonumber\\
P_n(x)  &= \frac{2n-1}{n}\frac{2\vartheta}{\Delta \vartheta}\: P_{n-1}(x)
-\frac{2n-1}{n}\frac{1}{B}\:P_{n-1}(x)\nonumber \\ &-\frac{n-1}{n}\: P_{n-2}(x)\;,
\end{align} where $B=\Delta\vartheta/(2\bar{\vartheta})=(\vartheta_{\mathrm{min}}-\vartheta_{\mathrm{max}})/(\vartheta_{\mathrm{min}}+\vartheta_{\mathrm{max}})$ is the relative interval width. Note that the first term in the recursive relation has the highest polynomial order, and the rest have lower order, respectively. We write the Legendre polynomials as polynomials in $\vartheta$

\begin{equation}
P_n(x)=\sum_{i=0}^{n}C_{ni}\:\vartheta^i\;,
\end{equation}
with $C_{ni}$ coefficients

\begin{align}
 C_{00} & =1\;,\nonumber \\
 C_{11} & =\frac{2}{\Delta \vartheta}\;,\nonumber\\ C_{10} & =-\frac{1}{B}\;, \nonumber\\
 C_{ni} & =\frac{2n-1}{n}\frac{2}{\Delta \vartheta}\: C_{n-1,i}
-\frac{2n-1}{n}\frac{1}{B}\:C_{n-1,i-1}\nonumber \\&-\frac{n-1}{n}\: C_{n-2,i}\;,
\end{align} where $C_{ni}=0$ if $i>n$ or $i<0$. The $a_{ni}$ coefficients are

\begin{equation}
a_{ni}=\sqrt{\frac{2n+3}{2}}\:C_{n+1,i}\;.
\end{equation}

The $W_n^\mathrm{Lin}$ weight functions computed the semi-analytic way
are considerably less computationally demanding, although, formulae
$\eqref{Snrec}$ and $\eqref{Snclosed}$ are not stable as far as we
investigated. The $S_n$ functions blow up for small arguments were
they should reach zero and show a noisy behavior. In addition, the
argument for which $S_n$ starts to behave as it should grows with $n$,
hence the resulting $W_n^\mathrm{Lin}$ functions become less and less
reliable for larger subscripts. However, that does not render them
useless since calculating the $W_n^\mathrm{Lin}$ functions from their
original integral form is very time consuming for larger arguments
were the $S_n$ functions become reliable. Since apart from $S_{-1}$
which needs to be stored once and can be loaded for further use, the
time taken to evaluate the rest of the $S_n$ functions does not depend
on their argument, which means one can in principle go to arbitrarily
high $\ell$-modes to calculate $W_n^\mathrm{Lin}(\ell)$.

Nevertheless, in practice one does not need to go to very high $\ell$-modes to evaluate the integral in \Eqt$\eqref{EnofWn}$, and find the E-mode COSEBIs; since as it is evident in \fig$\ref{Wn-plot}$, the $W_n$ functions die out rapidly at large $\ell$. By inspection of the plots, we deduced that the ratio of the largest peak (global maximum) and a peak at $\ell \approx 100\pi n/\vartheta_\mathrm{max}$ is of around 3-4 orders of magnitude. This property of the weight functions makes the infinite upper limit of \Eqt$\eqref{EnofWn}$ in practice manageable, i.e., the effective limits of the integral become finite, although they also depend on the shape of the power spectrum. In the present work the integrals involving $W_n$ functions are evaluated for a finite range of $\ell_\mathrm{min}=1$ to $\ell_\mathrm{max}\approx 100\pi n_{\mathrm{max}}/\vartheta_\mathrm{max}$, where $n_{\mathrm{max}}$ is the maximum number of modes considered in the analysis of the interest.

In the piece-wise method for calculating $W_n$, a Gaussian integration method (gauleg) of numerical recipes, \cite{numerical}, is used for each interval considered. The results of these integrals are summed up as the final result. There is a routine in the code which finds the consecutive minima and maxima of the zeroth order Bessel function, and puts them as the integration limits of the pieces. However, the low-$\ell$ values of the functions are not calculated in the same way, since in those regimes the oscillations of $T_n$ is more important compared to $\mathrm{J}_0$, and they cause numerical artifact. Instead one Gaussian integration method with higher accuracy parameter is used to evaluate them. The limit to change from one routine to the other is set by $\ell_{\mathrm{thresh}}\approx \pi n/ \vartheta_{\mathrm{max}}$ parameter.

\bibliographystyle{aa}
\bibliography{COSEBIs}
\end{document}